\documentclass[12pt]{iopart}

\usepackage{iopams} 
\usepackage{bm}
\usepackage{here}
\usepackage{tikz}
\usepackage{xcolor}
 \usepackage{circuitikzgit}
 \usepackage{pdflscape}
 \pgfcircdeclarebipolescaled{instruments}
{
    \anchor{text}{\pgfextracty{\pgf@circ@res@up}{\northeast}
        \pgfpoint{-.5\wd\pgfnodeparttextbox}{
            \dimexpr.5\dp\pgfnodeparttextbox+.5\ht\pgfnodeparttextbox+\pgf@circ@res@up\relax
        }
    }
}
{\ctikzvalof{bipoles/oscope/height}}
{josephson}
{\ctikzvalof{bipoles/oscope/height}}
{\ctikzvalof{bipoles/oscope/width}}
{
    \pgf@circ@setlinewidth{bipoles}{\pgfstartlinewidth}
    \pgfextracty{\pgf@circ@res@up}{\northeast}
    \pgfextractx{\pgf@circ@res@right}{\northeast}
    \pgfextractx{\pgf@circ@res@left}{\southwest}
    \pgfextracty{\pgf@circ@res@down}{\southwest}
    \pgfmathsetlength{\pgf@circ@res@step}{0.25*\pgf@circ@res@up}
    \pgfscope
        \pgfpathrectanglecorners{\pgfpoint{\pgf@circ@res@left}{\pgf@circ@res@down}}{\pgfpoint{\pgf@circ@res@right}{\pgf@circ@res@up}}
        \pgf@circ@draworfill
    \endpgfscope
    \pgfscope
      \pgfpathmoveto{\pgfpoint{\pgf@circ@res@left}{\pgf@circ@res@up}}%
      \pgfpathlineto{\pgfpoint{\pgf@circ@res@right}{\pgf@circ@res@down}}%
      \pgfpathmoveto{\pgfpoint{\pgf@circ@res@right}{\pgf@circ@res@up}}%
      \pgfpathlineto{\pgfpoint{\pgf@circ@res@left}{\pgf@circ@res@down}}%
      \pgfusepath{draw}
    \endpgfscope
}
\def\pgf@circ@josephson@path#1{\pgf@circ@bipole@path{josephson}{#1}}
\tikzset{josephson/.style = {\circuitikzbasekey, /tikz/to path=\pgf@circ@josephson@path, l=#1}}

\begin{document}



\title[Neglected $U(1)$ phase in the Schr\"{o}dinger representation of quantum mechanics]{Neglected $U(1)$ phase in the Schr\"{o}dinger representation of quantum mechanics and particle number conserving formalisms for superconductivity}

\author{Hiroyasu Koizumi}

\address{Division of Quantum Condensed Matter Physics, Center for Computational Sciences, University of Tsukuba, Tsukuba, Ibaraki 305-8577, Japan}
\ead{koizumi.hiroyasu.fn@u.tsukuba.ac.jp}
\vspace{10pt}

\begin{abstract}
Superconductivity is reformulated as a phenomenon in which a stable velocity field is created by a $U(1)$ phase neglected by Dirac in the Schr\"{o}dinger representation of quantum mechanics. The neglected phase gives rise to a $U(1)$ gauge field expressed as the Berry connection from many-body wave functions.
The inclusion of this gauge field transforms the standard particle-number non-conserving formalism of superconductivity to a particle-number conserving one
with many results of the former unaltered.
In other words, the new formalism indicates that the current standard one is an approximation that effectively takes into account this neglected $U(1)$ gauge field by employing the particle-number non-conserving formalism.
Since the standard and new formalisms are physically different,
conflicting results are predicted in some cases. We reexamine the Josephson relation and show that a capacitance contribution of the Josephson junction to the $U(1)$ phase is missing in the standard formalism, and inclusion of it indicates that the standard theory actually does not agree with the experiment while the new one does.
It is also shown that the dissipative quantum phase transition in Josephson junctions predicted in the standard theory does not exist in the new one in accordance with a recent experiment [A. Murani et al., Phys. Rev. X 10 (2020) 021003].
\end{abstract}


%

\section{Introduction}

After the success of the BCS theory for superconductivity \cite{BCS1957}, it is now widely-believed that 
superconductivity is a phenomenon of 
gauge symmetry breaking. The order parameter for the superconducting phase is a macroscopic wave function given by the expectation value of the electron-pair field operator \cite{Anderson}; in order to have non-zero expectation values for it, a particle-number non-conserving formalism is required. 
The original intention for the use of this formalism is to facilitate calculations involving electron pair-correlation that yields an energy gap \cite{BCS1957}. 
It is very practical for this purpose since it yields the total energy with an estimated relative error proportional to $N^{-1/2}$, where $N$ is the total number of the particles in the system; thus the error will be negligibly small compared with the energy itself for a bulk system with very large $N$ \cite{Peierls1991}.

Remarkably, the BCS theory successfully explains a number of pairing energy gap related effects; especially, it provides a way to calculate the superconducting transition temperature $T_c$ as the pairing energy gap formation temperature.
Surprisingly, a $U(1)$ phase arising in the particle-number non-conserving formalism turned 
out to be the essential ingredient for supercurrent generation. It brings about
the supercurrent related effects \cite{Josephson62}, which were predicted by Josephson using the Bogoliubov's particle-number non-conserving formalism of the BCS theory \cite{Bogoliubov1958}; and the predicted effects were confirmed by experiments \cite{Anderson-Rowell,Shapiro63}.

In spite of all those successes, the use of the particle-number non-conserving formalism is uncomfortable since it is sensible to consider that superconductivity also occurs in isolated systems with fixed number of particles.
Some researchers put significant efforts to restore the particle-number conservation in the formalism for superconductivity \cite{LeggettBook}. Such efforts are based on the idea of the Bose-Einstein condensation of electron pairs. However, it is not applicable for cases where the spatial overlapping of electron-pairs is significant
since it is not legitimate to treat electron-pairs as bosons in such cases. Thus, the Bose-Einstein condensation based refinement will not be able to explain the whole variety of superconductors.

The discovery of high transition temperature superconductors in cuprates upset the comfortable agreement between theory and experiment for superconductivity \cite{Muller1986}. At present, no widely accepted theory exists for this ``cuprate superconductivity''. The electron pairing gap with $d$-wave symmetry exists; however, $T_c$ is not identified as this gap formation temperature. The experiment indicates that $T_c$ corresponds to the stabilization temperature for nano-sized loop currents \cite{Kivelson95}; this view is supported by
the fact that the temperature dependence of the specific heat near $T_c$ resembles that of the three dimensional $XY$ model \cite{3Dxy1994}. Besides, there are a number of phenomena that indicate the
presence of loop currents in the pseudogap and superconducting phases \cite{Kerr1,Bourges2010,Varm2010,Hidekata2011,Nernst,Nernst2005,Nernst2}.

It is also notable that the normal state of the cuprate from which the superconducting state appears is not a simple metal;
the strong electron-lattice interaction causes the small lattice polaron formation and its relevance to superconductivity has been
experimentally verified \cite{Bianconi}; the strong electron repulsion in this system makes the 
parent compound an antiferromagnetic insulator \cite{AndersonBook,NeutronRev}, and the origin of the $d$-wave gap seems to be this strong repulsion.
Peculiarly enough, in spite of all those strong electron-lattice and electron-electron interactions expected to increase the effective mass,
the mass for the supercurrent carrier is that of free electron one according to the London moment measurement \cite{Verheijen1990}. Actually, this sharply contradicts the BCS theory since it predicts that this mass should be the effective mass of the normal metallic phase.
Furthermore, this London moment mass discrepancy problem is not just for the cuprate superconductivity, but also for all other superconductors; the free electron mass is observed universally in the London moment experiments \cite{Hirsch2013b}.

Recently, it is argued that the $U(1)$ phase that
appears in the particle-number non-conserving formalism is actually 
the $U(1)$ phase neglected by Dirac in the Schr\"{o}dinger's representation of quantum mechanics \cite{DiracSec22}. In this new theory, the mass in the London moment is the free electron mass, and the $U(1)$ phase variable for superconductivity appears with keeping the number of particles fixed \cite{koizumi2022}.

Let us explain this new theory succinctly, below, since the rest of this work depends on it.
In the Schr\"{o}dinger's representation, the momenta $p_r$'s conjugate to canonical coordinates $q_r$'s are given by
\begin{eqnarray}
p_r=-i \hbar {{\partial} \over {\partial q_r}}, \quad r=1, \cdots, n
\label{pr1}
\end{eqnarray}
where $n$ is the degrees of freedom in the many-body system.
According to the Heisenberg formulation of quantum theory \cite{Born-Jordan}, commutation relations
\begin{eqnarray}
[q_r, q_s]=0, \quad [p_r, p_s]=0, \quad  [q_r, p_s]=i \hbar \delta_{rs}
\end{eqnarray}
are more fundamental than the derivative representation of the momenta.
Actually, the following  $p_r$'s are also legitimate
\begin{eqnarray}
p_r=-i \hbar {{\partial} \over {\partial q_r}}+{{\partial F} \over {\partial q_r}}, \quad r=1, \cdots, n
\label{pr2}
\end{eqnarray}
where $F$ is a function of $q_1, \cdots, q_n$, and also of time $t$. 
Dirac argued that $F$ can be removed by
the following transformation of the wave function,
\begin{eqnarray}
\psi(q_1, \cdots, q_n) \rightarrow e^{i \gamma} \psi(q_1, \cdots, q_n) 
\label{gamma}
\end{eqnarray}
where $\gamma$ is related to $F$ by
$
F=\hbar \gamma + \mbox{constant}
$,
thus, we can always use $p_r$'s in Eq.~(\ref{pr1}) \cite{DiracSec22}.
The standard calculation obtains $e^{i \gamma} \psi(q_1, \cdots, q_n)$ as a whole by employing a finite number of basis functions, thus, $e^{i \gamma}$ is not considered explicitly.
Contrary to the standard procedure, however, there are cases where $e^{i \gamma}$ should be considered as an
extra degree-of-freedom \cite{koizumi2022}. 
Actually, 
the condition $[p_r, p_s]=0$ with Eq.~(\ref{pr2}) requires the following condition
\begin{eqnarray}
{\partial \over {\partial q_s}}{{\partial F} \over {\partial q_r}}-{\partial \over {\partial q_r}}{{\partial F} \over {\partial q_s}}=0
\label{eqSingular}
\end{eqnarray}
It can be violated at points where the amplitude of $\psi(q_1, \cdots, q_n)$ is zero, and topological defects may exist there. In such cases, $\gamma$ is expressed as a nontrivial Berry connection from many-body wave functions given by
\begin{eqnarray}
\gamma= \sum_{j=1}^{N} \int_0^{{\bf r}_j} {\bf A}^{\rm MB}_{\Psi}({\bf r}', t) \cdot d{\bf r}' 
\label{gamma1}
\end{eqnarray}
where ${\bf A}^{\rm MB}_{\Psi}({\bf r},t)$ is
 \begin{eqnarray}
{\bf A}^{\rm MB}_{\Psi}({\bf r},t)=
{ 1\over {\hbar \rho({\bf r},t)}} {\rm Re} \left\{
 \int d\sigma_1  d{\bf x}_{2}  \cdots d{\bf x}_{N}
 \Psi^{\ast}({\bf r}, \sigma_1, \cdots, {\bf x}_{N}, t)
  (-i \hbar \nabla_{\bf r} )
\Psi({\bf r}, \sigma_1, \cdots, {\bf x}_{N}, t) \right\}, 
\label{Afic}
\nonumber
\\
\label{velocityA}
\end{eqnarray}
`$\rm{Re}$' here denotes the real part, $\Psi$ is the total wave function, ${\bf x}_i$ collectively stands for the coordinate ${\bf r}_i$ and the spin $\sigma_i$ of the $i$th electron; $t$ is time; $-i \hbar \nabla_{\bf r}$ is the Schr\"{o}dinger's momentum operator for the coordinate vector ${\bf r}$, and $\rho({\bf r},t)$ is the number
density calculated from $\Psi$. 
This Berry connection is obtained by regarding ${\bf r}$ as the ``adiabatic parameter''\cite{Berry}; it detects topological singularities of the connection (points of the violation of Eq.~(\ref{eqSingular})); the adiabatic assumption of the parameter is disregarded.
Note that Dirac considered
such singularities in relation to the magnetic monopole before Berry \cite{Monopole}. 
Note also that ${\bf A}^{\rm MB}_{\Psi}({\bf r},t)$ is a many-body quantity since its value for particle labeled `$1$' at ${\bf r}_1={\bf r}$ arises from variables for other particles; thus, it is absent for the hydrogen problem solved by Schr\"{o}dinger \cite{Schrodinger}.
The connection ${\bf A}^{\rm MB}_{\Psi}({\bf r},t)$ generates a topologically protected velocity field (that corresponds to ${{\partial F} \over {\partial q_r}}$ in Eq.~(\ref{pr2})) and the supercurrent is generated by it \cite{koizumi2022}.

The new superconductivity theory quantizes the collective mode generated by ${\bf A}^{\rm MB}_{\Psi}$. Then, the fluctuation of the number of electrons participating in this mode is taken into account; actually, this fluctuation lowers the total energy by the electron-pair formation.
The new theory exhibits the same Bogoliubov excitation spectrum to the standard theory, thus, it
captures many of the important results obtained by the standard one. 
Besides, it removes some of the shortcomings of the standard theory, such as the mass discrepancy problem in the London moment \cite{koizumi2020b,koizumi2020c,koizumi2022}. 
In the new theory, the electron-pair formation is not the essential ingredient for the supercurrent generation; 
its role is to stabilize the velocity field of the supercurrent.
Thus, if the stabilization of the supercurrent and the appearance of the pairing energy gap
occur simultaneously, $T_c$ is given by the pairing gap formation temperatures, as is seen in many superconductors; however, it does not hold in the cuprate superconductivity.

In the present work, we examine the new theory of superconductivity 
putting emphasis on Josephson effects related phenomena since the standard and new theories predict  conflicting results in them. We also include some materials previously published in a succinct form 
so that the arguments can be followed without consulting previous publications.
Most notable conclusions in the present work are the following two: 1) a capacitance contribution of the Josephson junction to the $U(1)$ phase is missing in the standard theory, and inclusion of it indicates that the standard one actually does not agree with the experiment while the new one does; 2) 
the dissipative quantum phase transition in Josephson junctions predicted in the standard theory does not exist in the new one in accordance with an experiment \cite{PhysRevX2021a}.

The organization of the present work is as follows: In Section~\ref{sec1}, we explain so-called ``Bloch's theorem'' for the impossibility of a current carrying  ground state \cite{Bohm-Bloch}, and how it is overcome in the new theory. In Section~\ref{Sec2},
we revisit the flux quantization and Josephson effects. In Section~\ref{Sec3}, we 
examine the Josephson relation and the Ambegaokar-Baratoff relation \cite{Ambegaokar} from the view point of the new theory. In Section~\ref{Sec4},
a practical way to include the Dirac's neglected $U(1)$ phase in the calculation of superconducting electronic states is explained. In Section~\ref{Sec5},
the absence of the dissipative quantum phase transition in Josephson junctions \cite{Schmid1983,PhysRevX2021b,PhysRevX2021a,PhysRevX2021c} is discussed.
 In Section~\ref{Sec6}, we conclude the present work by mentioning implications of the present new theory to the cuprate superconductivity and the superfluidity of bosons.

%

\section{ ``Bloch's theorem'' for the impossibility of a current carrying  ground state and its violation by the neglected $U(1)$ phase collective mode}
\label{sec1}

First, we revisit so-called ``Bloch's theorem'' that states `the current carrying  ground state
is impossible' \cite{Bohm-Bloch}.
By following Bohm \cite{Bohm-Bloch}, we consider the $N$-electron system with the Hamiltonian
\begin{eqnarray}
H=\sum_{j=1}^N \left[-{ \hbar^2 \over {2 m_e}}\nabla_j^2 +U({\bf r}_j)\right]+{ 1 \over 2}\sum_{i\neq j} V({\bf r}_i, {\bf r}_j),
\label{bloch-15}
\end{eqnarray}
where $U({\bf r}_j)$ is a single-particle potential energy, and $V({\bf r}_i, {\bf r}_j)$ is a two-particle interaction energy. Here, the Schr\"{o}dinger's momenta $-i\hbar \nabla_j$ are used.
The ground state wave function is given by
\begin{eqnarray}
\psi_{\rm G}({\bf x}_1, \cdots, {\bf x}_N)
\end{eqnarray}
According to the ``Bloch's theorem'', this state is currentless. We will prove this using proof by contradiction as has been done by Bohm, below.

Let us denote the total momentum of $\psi_{\rm G}$ by ${\bf P}_0$; then the current density is given by
\begin{eqnarray}
{\bf j}_0=-{e \over m_e}{\bf P}_0 
\end{eqnarray}
We modify $\psi_{\rm G}$ to $\psi_{\rm Bohm}$ by changing the momentum of each electron by $\delta {\bf P}$,
\begin{eqnarray}
\psi_{\rm Bohm}=e^{{i \over \hbar} \delta {\bf P}\cdot \sum_{j=1}^N {\bf r}_j}\psi_{\rm G}({\bf x}_1, \cdots, {\bf x}_N)
\label{eqphi}
\end{eqnarray}
The total momentum for $\psi_{\rm Bohm}$ is given by 
\begin{eqnarray}
{\bf P}={\bf P}_0+N\delta {\bf P} 
\end{eqnarray}
The potential energy part does not change, thus, the difference of the total energies of the two states only comes from the kinetic energies. Let us denote the kinetic energies for $\psi_{\rm Bohm}$ and $\psi_{\rm G}$ by $T$ and $T_0$, respectively.
Their difference is given by
\begin{eqnarray}
T-T_0&=&{N \over m_e}{\bf P}_0 \cdot \delta {\bf P} +{N \over {2m_e}}(\delta {\bf P})^2
\nonumber
\\
&=&-{N \over e}{\bf j}_0 \cdot \delta {\bf P} +{N \over {2m_e}}(\delta {\bf P})^2
\label{eqT0}
\end{eqnarray}
If ${\bf j}_0$ is not zero, we can make this difference negative by
suitably choosing $\delta {\bf P}$. This means that the lower energy state than the ground state is realized
if ${\bf j}_0 \neq0$. This is a contradiction, thus, ${\bf j}_0 \neq0$ is false and ${\bf j}_0=0$ should hold.

In the new theory, the origin of the superconducting state is the presence of non-trivial neglected $U(1)$ phase.
It gives rise to a collective mode, and the fluctuation of number of particles participating 
in it occurs. 
Although the kinetic energy is increased by the presence of this collective mode, it can lower the potential energy by suitably choosing the fluctuation of the participating particle number as will be explained later.
Actually, if the potential energy decreases is larger than the kinetic energy increase,
the current carrying ground state is realized.

Now we shall explain the supercurrent generation by the collective mode. The wave function corresponding to $\psi_{\rm Bohm}$ in Eq.~(\ref{eqphi}) is $\Psi$ in Eq.~(\ref{Afic}).
 Let us calculate the Berry connection from $\Psi$; it is given by
 \begin{eqnarray}
{\bf A}^{\rm MB}_{\Psi}({\bf r},t)=-i \langle n_{\Psi}({\bf r},t) |\nabla_{\bf r}  |n_{\Psi}({\bf r},t) \rangle
\end{eqnarray}
where $|n_{\Psi}({\bf r},t) \rangle$ is a ket vector with parameter ${\bf r}$ defined by
\begin{eqnarray}
\langle  \sigma_1, {\bf x}_{2}, \cdots, {\bf x}_{N_e} |n_{\Psi}({\bf r},t) \rangle = { {\Psi({\bf r}, \sigma_1, {\bf x}_{2}, \cdots, {\bf x}_{N_e},t)} \over {|C_{\Psi}({\bf r} ,t)|^{{1 \over 2}}}}
\end{eqnarray}
Here $|C_{\Psi}({\bf r} ,t)|$ is a normalization constant given by
 \begin{eqnarray}
|C_{\Psi}({\bf r} ,t)|=\int d\sigma_1 d{\bf x}_{2} \cdots d{\bf x}_{N_e}\Psi({\bf r},\sigma_1, {\bf x}_{2}, \cdots)\Psi^{\ast}(\sigma_1, {\bf x}, {\bf x}_{2}, \cdots)
\end{eqnarray}
which yields the normalized ket vector, satisfying $\langle n_{\Psi}({\bf r},t) |n_{\Psi}({\bf r},t) \rangle=1$.
The explicit expression for ${\bf A}^{\rm MB}_{\Psi}({\bf r},t)$ is given already in Eq.~(\ref{velocityA}),
where $\rho$ in it is given by
 \begin{eqnarray}
\rho({\bf r},t)=\int d\sigma_1 d{\bf x}_{2} \cdots d{\bf x}_{N}\Psi({\bf r},\sigma_1, {\bf x}_{2}, \cdots, {\bf x}_{N},t)\Psi^{\ast}({\bf r}, \sigma_1, {\bf x}_{2}, \cdots, {\bf x}_{N},t)
\nonumber
\\
\end{eqnarray}
The equation (\ref{velocityA}) indicates that ${\bf A}^{\rm MB}_{\Psi}({\bf r},t)$
is a velocity field multiplied by $m_e/\hbar$ with ${\bf p}_{\bf r}=-i \hbar \nabla_{\bf r}$.

Now we consider a wave function corresponding to $\psi_{\rm G}$ in Eq.~(\ref{eqphi}). We denote it by $\Psi_0$,
\begin{eqnarray}
\Psi_0 ({\bf x}_1, \cdots, {\bf x}_{N_e},t)=\Psi ({\bf x}_1, \cdots, {\bf x}_{N_e},t)\exp\left(- i \sum_{j=1}^{N_e} \int_{0}^{{\bf r}_j} {\bf A}_{\Psi}^{\rm MB}({\bf r}',t) \cdot d{\bf r}' \right)
\label{wavef0}
\nonumber
\\
\end{eqnarray}
Note that it is a currentless state since the Berry connection for it is zero,
\begin{eqnarray}
{\bf A}^{\rm MB}_{\Psi_0}({\bf r},t)=-i \langle n_{\Psi_0}({\bf r},t) |\nabla_{\bf r}  |n_{\Psi_0}({\bf r},t) \rangle =0, 
\end{eqnarray}
due to the cancelation of the Berry connection from $\Psi$ and that from the phase factor $\exp \left(  -{i \over 2} \sum_{j=1}^{N} \chi({\bf r}_j,t)\right)$.
Conversely, $\Psi$ is given by 
\begin{eqnarray}
\Psi=\exp\left( i \sum_{j=1}^{N_e} \int_{0}^{{\bf r}_j} {\bf A}_{\Psi}^{\rm MB}({\bf r}',t) \cdot d{\bf r}' \right)\Psi_0 
\end{eqnarray}
The comparison of the above $\Psi$ with the wave function in Eq.~(\ref{gamma}) indicates that $\gamma$ is given by
\begin{eqnarray}
\gamma({\bf r}_1, \cdots,{\bf r}_N,t)= \sum_{j=1}^{N} \int_0^{{\bf r}_j} {\bf A}^{\rm MB}_{\Psi}({\bf r}',t) \cdot d{\bf r}' 
\end{eqnarray}
For convenience sake, we use the following $\chi$,
 \begin{eqnarray}
{ {\chi({\bf r},t)}}= - 2\int^{{\bf r}}_0 {\bf A}_{\Psi}^{\rm MB}({\bf r}',t) \cdot d{\bf r}' 
\end{eqnarray}
Then, $\gamma$ is expressed as
\begin{eqnarray}
\gamma= -{1 \over 2}\sum_{j=1}^{N} \chi({\bf r}_j,t)
\label{gamma2}
\end{eqnarray}
and
 $\Psi$ is given by
\begin{eqnarray}
\Psi({\bf x}_1, \cdots, {\bf x}_N,t)=\exp \left(  -{i \over 2} \sum_{j=1}^{N} \chi({\bf r}_j,t)\right)
\Psi_0({\bf x}_1, \cdots, {\bf x}_N,t)
\label{single-valued}
\end{eqnarray}
Due to the presence of the phase factor, the total energy increases; it is given by
\begin{eqnarray}
E[\chi]-E_0={\hbar^2 \over {8 m_e}}\int d^3r \rho({\bf r}, t) \left[\nabla \chi ({\bf r},t)\right]^2
\label{eqEchi}
\end{eqnarray}
where $E[\chi]$ is the total energy for $\Psi$ which is a functional of $\chi$, and $E_0$ is the total energy for $\Psi_0$.
This should be compared to Eq.~(\ref{eqT0}) with taking into account the fact that $\Psi_0$ is currentless.

Let us denote the velocity field for $\Psi$ by {\bf v}. It is given by
\begin{eqnarray}
{\bf v}
={\hbar \over {2m_e}} \nabla \chi 
\label{eq12}
\end{eqnarray}
The supercurrent is generated by this $\nabla \chi$ due to the topological protection
as will be explained below.
The topological protection here means the realization of the situation where
\begin{eqnarray}
 w_C[\chi]={ 1 \over {2 \pi}} \oint_C \nabla \chi \cdot d{\bf r}
 \label{eqwindingN}
\end{eqnarray}
is time-independent 
\begin{eqnarray}
{d \over {dt}} w_C[\chi]=0
\label{eq15}
\end{eqnarray}
Here, $C$ is a loop in the coordinate space of the system, and non-zero $w_C[\chi]$ occurs when singularities of $\chi$ exist within it; $w_C[\chi]$ is an integer called the winding number. The time-derivative of the velocity field in normal metals is often expressed using the relaxation time approximation 
\begin{eqnarray}
{{d{\bf v}} \over {dt}}=-{1 \over \tau}{\bf v}
\label{eq16}
\end{eqnarray}
where $\tau$ is the relaxation time.
From  Eqs.~(\ref{eq12}) and (\ref{eq15}), the following relation is obtained
\begin{eqnarray}
\tau {d \over {dt}}w_C[\chi] =-w_C[\chi]
\label{eq17}
\end{eqnarray}
The condition in Eq.~(\ref{eq15}) with $w_C[\chi] \neq 0$ requires
$\tau$ to be $\infty$. This also means that the current produced by $\nabla \chi$ with $w_C[\chi] \neq 0$ is a non-decaying supercurrent.

The non-zero winding number $\chi$ arises when electrons perform spin-twisting itinerant motion.
For example, if the electron motion is such that it is a circular motion along loop $C$ with spin-twisting once around; then, the spinor character of the spin function gives rise to the sign-change of $\Psi_0$, i.e., it changes sign when an excursion is performed along $C$.
In order to have the single-valued $\Psi$, the phase factor part must make a compensating sign-change for the same encircling along $C$. This compensating sign-change is achieved when odd-integral $w_C[\chi]$.

For the realization of the ground state with this supercurrent, the kinetic energy increase caused by $\nabla \chi$ must be compensated  by the decrease of the potential energy caused by the fluctuation of the number of particles participating in the collective mode of $\chi$.
In the standard BCS theory, the potential energy decreases is achieved by electron-pairing, which is taken into account using the particle-number fluctuating ground state.
This ``particle-number fluctuation'' 
is replaced by the fluctuation of number of particles participating in the collective mode $\chi$
in the new theory.

Let us examine the reduction of the potential energy by starting with the BCS theory. It uses the following variational state vector to calculate the potential energy decrease
\begin{eqnarray}
|{\rm BCS}\rangle =
\prod_{\bf k}(u_{\bf k}+e^{i\theta}v_{\bf k}c^{\dagger}_{{\bf k} \uparrow} c^{\dagger}_{-{\bf k} \downarrow} )|{\rm vac} \rangle
\label{theta1}
\end{eqnarray}
where $|{\rm vac} \rangle$ is the vacuum state that satisfies
\begin{eqnarray}
c_{{\bf k} \sigma}|{\rm vac} \rangle=0
\end{eqnarray}
with $c_{{\bf k} \sigma}$ being the annihilation operator for the conduction electron with effective mass $m^{\ast}$, the wave vector ${\bf k}$, and spin $\sigma$.
The creation operator for the same electron is given by $c^{\dagger}_{{\bf k} \sigma}$, and $u_{\bf k}$ and $v_{\bf k}$ are variational parameters \cite{BCS1957}. The phase $\theta$ appears here as a meaningful parameter due to the ``particle-number fluctuation''.
 The spatial variation of $\theta$ gives rise to the Nambu-Goldstone mode \cite{Nambu1960}, which is  an essential
 ingredient for the supercurrent generation. 
Another way to look at the BCS state in Eq.~(\ref{theta1}) is to consider it as a vacuum of the following Bogoliubov operators 
\cite{Bogoliubov58},
\begin{eqnarray}
\gamma_{{\bf k} 0}^{\rm BCS} &= &u_{\bf k} e^{-{i \over 2} \theta}  c_{{\bf k} \uparrow}-v_{\bf k} e^{{i \over 2} \theta}  c_{-{\bf k} \uparrow}^{\dagger}
\nonumber
\\
\gamma_{-{\bf k} 1} ^{\rm BCS}&=&u_{\bf k} e^{-{i \over 2} \theta}  c_{-{\bf k} \downarrow}+v_{\bf k} e^{{i \over 2} \theta}  c_{{\bf k} \uparrow}^{\dagger}
\end{eqnarray}
It satisfies 
\begin{eqnarray}
\gamma_{{\bf k} 0}^{\rm BCS}|{\rm BCS}\rangle =0, \quad  \gamma_{-{\bf k} 1}^{\rm BCS}|{\rm BCS}\rangle =0 
\end{eqnarray}
The BCS result is interpreted that the potential energy of the ground state is lowered by the electron-pair formation,
and excitations from it are the Bogoliubov excitations generated by $(\gamma_{{\bf k} 0}^{\rm BCS})^{\dagger}$ and $(\gamma_{-{\bf k} 1}^{\rm BCS})^{\dagger}$ \cite{deGennes}.

Now we consider the new theory. We use a similar state to $|{\rm BCS}\rangle$ with conserving the particle number.
For this purpose, we utilize the collective mode described by $\chi$. We consider the fluctuation of the number of particles participating in this collective mode with keeping the particle number fixed.
In order to include this fluctuation, we quantize the $\chi$ mode as a boson field just like the photon field.
The quantization is achieved by using the following Lagrangian 
\begin{eqnarray}
L&=&\langle \Psi | i\hbar {\partial \over {\partial t}} -H |\Psi \rangle
\nonumber
\\
&=&\int d^3 r \hbar { \dot{\chi} \over 2}\rho+ i\hbar \langle \Psi_0 |{\partial \over {\partial t}}|\Psi_0 \rangle-\langle \Psi | H |\Psi \rangle
\end{eqnarray}
The conjugate field to $\chi$, $\pi_{\chi}$, given by
\begin{eqnarray}
\pi_{\chi}={{\delta L} \over {\delta \dot{\chi} }}={ \hbar \over 2} \rho
\end{eqnarray}
The canonical quantization conndition $[\chi({\bf r},t),\pi_{\chi} ({\bf r}',t)]=i\hbar \delta({\bf r}-{\bf r}')$
yields the following relation
\begin{eqnarray}
[\chi({\bf r},t),\rho({\bf r}',t)]=2i\delta({\bf r}-{\bf r}')
\label{eq16-4-68}
\end{eqnarray}
We consider the case where $\chi$ and $\rho$ are time-independent in the following. 
From $\chi({\bf r})$ and $\rho({\bf r})$, we construct the following boson field operators 
\begin{eqnarray}
\psi_{\chi}^{\dagger}({\bf r})=\sqrt{\rho({\bf r})}e^{{ i \over 2}
\chi({\bf r})}, \quad \psi_{\chi}({\bf r})=e^{-{ i \over 2}\chi({\bf r})}\sqrt{\rho({\bf r})}
\end{eqnarray}
They satisfy the following commutation relation
\begin{eqnarray}
[ \psi_{\chi}({\bf r}), \psi_{\chi}^{\dagger}({\bf r}')]=\delta({\bf r}-{\bf r}')
\label{compsi}
\end{eqnarray}
We further define the following boson operators $B^{\dagger}_{\chi}$ and $B_{\chi}$
\begin{eqnarray}
B^{\dagger}_{\chi} =\int d^3r  \psi_{\chi}^{\dagger}({\bf r}),
\quad B_{\chi} =\int d^3r \psi_{\chi}({\bf r})
\end{eqnarray}
that satisfy the commutation relation
\begin{eqnarray}
[ B_{\chi}, B_{\chi}^{\dagger}]=1
\label{eqBoson}
\end{eqnarray}
Then, the following number operator can be constructed
\begin{eqnarray}
\hat{N}_{\chi}=B_{\chi}^{\dagger} B_{\chi}
\end{eqnarray}

The commutation relation in Eq.~(\ref{eqBoson}) is the one for the harmonic oscillator, and the eigenstate $|{N}_{\chi} \rangle$ 
for the number operator satisfies
\begin{eqnarray}
\hat{N}_{\chi}|{N}_{\chi} \rangle
={N}_{\chi}|{N}_{\chi} \rangle
\end{eqnarray}
We identify ${N}_{\chi}$ as the number of electrons participating in the collective mode of $\chi$.
The conjugate operator to $\hat{N}_{\chi}$, $\hat{X}$, is defined through the following relation
\begin{eqnarray}
 B_{\chi}^{\dagger} =\sqrt{\hat{N}_{\chi}}e^{i{1 \over 2}\hat{X}}, \quad B_{\chi} =e^{-i{1 \over 2}\hat{X}}
\sqrt{\hat{N}_{\chi}}
\end{eqnarray}
They satisfy the commutation relation 
\begin{eqnarray}
[\hat{N}_{\chi}, e^{\pm i {1 \over 2}\hat{X}}] =\pm e^{\pm i {1 \over 2}\hat{X}}
\end{eqnarray}
which yields
\begin{eqnarray}
e^{\pm i {1 \over 2}\hat{X}} |{N}_{\chi} \rangle \propto |{N}_{\chi} \pm 1\rangle
\label{eqNumberChanging}
\end{eqnarray}
The above relation indicates that $e^{ -i{1 \over 2} \hat{X}}$ is the number changing operator that decreases the
number of electrons participating in the collective mode by one, and $e^{i{1 \over 2}\hat{X}}$ increases by one. Especially, $e^{ -i \hat{X}}$ reduces the number by two.

In the new theory,
 $|{\rm BCS}\rangle$ is replaced 
by $|{\rm Gnd}\rangle$
\begin{eqnarray}
|{\rm Gnd}\rangle =
\prod_{\bf k}(u_{\bf k}+v_{\bf k} c^{\dagger}_{{\bf k} \uparrow} c^{\dagger}_{-{\bf k} \downarrow}e^{-i \hat{X}} )|{\rm Cnd} \rangle
\label{chi1}
\end{eqnarray}
where $|{\rm Cnd} \rangle$ is the state vector corresponding to $|{N}_{\chi}=N \rangle$, i.e., all the particles are participating in the collective mode.
The comparison of Eqs.~(\ref{theta1}) and (\ref{chi1}) indicates the following correspondence 
 \begin{eqnarray}
 e^{i\theta} \rightarrow e^{-i \hat{X}}, \quad |{\rm vac}\rangle \rightarrow |{\rm Cnd}  \rangle
\end{eqnarray}
between the BCS theory and new one. The same $u_{\bf k}$'s and $v_{\bf k}$'s appear.
The electron pair creation operators $e^{i\theta}c^{\dagger}_{{\bf k} \uparrow} c^{\dagger}_{-{\bf k} \downarrow}$ in $|{\rm BCS} \rangle$ 
is replaced by $c^{\dagger}_{{\bf k} \uparrow} c^{\dagger}_{-{\bf k} \downarrow}e^{-i \hat{X}}$ in $|{\rm Gnd} \rangle$. The latter term decreases the number of electrons participating in the collective mode by two and increases
 the number of electrons in single-particle motion states by two. Thus, the particle number is fixed. 
 We use the expression $|{\rm Gnd}(N) \rangle$ when we want to explicitly express the number of the particles in $|{\rm Gnd} \rangle$ in the following.
 
The Bogoliubov operators in the new theory are given by
\begin{eqnarray}
\gamma_{{\bf k} 0}&=&u_{\bf k} e^{{i \over 2} \hat{X}}  c_{{\bf k} \uparrow}-v_{\bf k} e^{-{i \over 2} \hat{X}}  c_{-{\bf k} \uparrow}^{\dagger}
\nonumber
\\
\gamma_{-{\bf k} 1}&=& u_{\bf k} e^{{i \over 2} \hat{X}}  c_{-{\bf k} \downarrow}+v_{\bf k} e^{-{i \over 2} \hat{X}}  c_{{\bf k} \uparrow}^{\dagger}
\end{eqnarray}
They conserve the particle number, and satisfy
\begin{eqnarray}
\gamma_{{\bf k} 0}|{\rm Gnd}\rangle =0, \quad  \gamma_{-{\bf k} 1}|{\rm Gnd}\rangle =0 
\end{eqnarray}
Due to these properties, the same energy decrease by the electron pairing is achieved, and the same Bogoliubov excitation spectrum is obtained \cite{koizumi2019}.
The ground state $|{\rm Gnd}\rangle$ is equipped with the velocity field in Eq.~(\ref{eq12}).  If it is non-zero, non-zero current exists; then, the ``Bloch's theorem'' is violated.

%

\section{Flux Quantization and Josephson effects}
\label{Sec2}

Now we consider the Josephson effects.
Let us consider a Josephson (superconductor-insulator-superconductor (SIS)) junction.
The hopping Hamiltonian between the two superconductors (called `left' and `right', respectively) across the insulator is given by
  \begin{eqnarray}
H_{LR}=-\sum_{\sigma}T_{LR}e^{i {e \over {\hbar c} } \int^{{\bf r}_R}_{{\bf r}_L} d{\bf r} \cdot {\bf A}^{\rm em}}
  c^{\dagger}_{L \sigma} c_{R \sigma}+{\rm h.c.}
\label{juncH}
\end{eqnarray}
where quantities for the left superconductor is labeled with $L$, and those for the right superconductor are labeled with $R$; $T_{LR}$ is the hoping matrix element of the junction, and  $c^{\dagger}_{L \sigma}$ and  $c_{R \sigma}$ denote
creation and annihilation operators for electrons with spin $\sigma$, respectively. The phase factor $e^{i {e \over {\hbar c} } \int^{{\bf r}_R}_{{\bf r}_L} d{\bf r} \cdot {\bf A}^{\rm em}}$ arises due to the magnetic field ${\bf B}^{\rm em}$, which is related to the vector potential ${\bf A}^{\rm em}$ as ${\bf B}^{\rm em}=\nabla \times {\bf A}^{\rm em}$; ${\rm h.c.}$ denotes the hermitian conjugate term. We take this Hamiltonian as a perturbation.

\subsection{BCS superconducting state case}

First, we consider the BCS superconducting state case.
The state vectors for the two superconductors are given by
\begin{eqnarray}
|{\rm BCS}_L\rangle &=&
\prod_{\bf k}(u_{L {\bf k}}+e^{i\theta_L} v_{L {\bf k}}c^{\dagger}_{L {\bf k} \uparrow} c^{\dagger}_{L -{\bf k} \downarrow} )|{\rm vac} \rangle
\nonumber
\\
|{\rm BCS}_R\rangle &=&
\prod_{\bf k}(u_{R {\bf k}}+e^{i\theta_R} v_{R {\bf k}}c^{\dagger}_{R {\bf k} \uparrow} c^{\dagger}_{R -{\bf k} \downarrow} )|{\rm vac} \rangle
\label{thetaLR}
\end{eqnarray}

The nonzero perturbation for tunneling arises from the second order in $H_{LR}$. In this case, the tunneling is due to the electron-pair transfer between the two superconductors. The supercurrent related terms contain the phase factor $e^{i\phi}$, where the angular variable $\phi$ is given by
\begin{eqnarray}
\phi=\theta_R-\theta_L+{{2e}\over {\hbar c}}
\int^{{\bf r}_R}_{{\bf r}_L}
 {\bf A}^{\rm em} \cdot d{\bf r}
 \label{eq:phi}
\end{eqnarray}
The second order perturbation of $H_{LR}$ yields ${{2e}\over {\hbar c}}
\int^{{\bf r}_R}_{{\bf r}_L}
 {\bf A}^{\rm em} \cdot d{\bf r}$, and $\theta_R-\theta_L$ arises from the
 phases $e^{i\theta_L}$ and $e^{i\theta_R}$ of the superconducting states $|{\rm BCS}_L\rangle$ and $|{\rm BCS}_R\rangle$, respectively.
The formula in Eq.~(\ref{eq:phi}) indicates that the following sum is the gauge invariant combination,
\begin{eqnarray}
\nabla \theta + {{2e}\over {\hbar c}}{\bf A}^{\rm em}
\end{eqnarray}
The time-component partner for the above combination is 
\begin{eqnarray}
\partial_t \theta - {{2e}\over {\hbar }}\phi^{\rm em}
 \label{eq:phi2}
\end{eqnarray}
where $\phi^{\rm em}$ is the scalar potential for the electromagnetic field.
The above is also gauge invariant.

Using Eqs.~(\ref{eq:phi})-(\ref{eq:phi2}), the time-derivative of $\phi$ is given by
\begin{eqnarray}
\dot{\phi}&=&\int^{{\bf r}_R}_{{\bf r}_L} \nabla \partial_t{\theta} \cdot d{\bf r}+{{2e}\over {\hbar c}}
\int^{{\bf r}_R}_{{\bf r}_L} \partial_t {\bf A}^{\rm em} \cdot d{\bf r}
\nonumber
\\
&=&\int^{{\bf r}_R}_{{\bf r}_L} \nabla \left(
\partial_t {\theta}-{{2e}\over {\hbar }}\phi^{\rm em} \right)
 \cdot d{\bf r}-{{2e}\over {\hbar }}
\int^{{\bf r}_R}_{{\bf r}_L} {\bf E}^{\rm em} \cdot d{\bf r}
\label{eq25}
\end{eqnarray}
where $ {\bf E}^{\rm em}$ is the electric field given by
\begin{eqnarray}
{\bf E}^{\rm em}=-{1 \over c} \partial_t{\bf A}^{\rm em}  -\nabla \phi^{\rm em}
\end{eqnarray}
This shows that the time derivative $\dot{\phi}$ is composed of two gauge invariant terms.
The first term in Eq.~(\ref{eq25}) arises from the chemical potential difference of the two superconductors, which is expressed as
\begin{eqnarray}
\int^{{\bf r}_R}_{{\bf r}_L} \nabla \left(
\partial_t {\theta}-{{2e}\over {\hbar }}\phi^{\rm em} \right)
 \cdot d{\bf r}={{2eV}\over {\hbar }}
 \label{eqV1}
\end{eqnarray}
where $V$ is the voltage across the junction.
From this term, Josephson obtained the Josephson relation \cite{Josephson62} given by
\begin{eqnarray}
\dot{\phi}={{2eV}\over {\hbar }}
\end{eqnarray}
The second term in Eq.~(\ref{eq25}) that contains ${\bf E}^{\rm em}$ is due 
to the presence of the electric field between the two superconductors. The presence of this term 
indicates that the Josepshon junction may contain a capacitor contribution.
This term is absent in the Josephson's derivation \cite{Josephson62}, and in the metal-superconductor tunneling calculation by Cohen et al. \cite{Cohen1962}.
For the latter case, the capacitance effect is negligible, thus, its omission is justifiable.
However, for the Josephson junction case, it is not usually negligible as is manifested in the use of the RCSJ model \cite{TinkhamText}. It is also an important ingredient of the superconducting qubit as will be seen in Section~\ref{Sec5}.
 
From the capacitance contribution, we have
\begin{eqnarray}
-{{2e}\over {\hbar }}
\int^{{\bf r}_R}_{{\bf r}_L} {\bf E}^{\rm em} \cdot d{\bf r}
={{2eV}\over {\hbar }}
 \label{eqV2}
\end{eqnarray}
due to the fact that the voltage across the capacitor part of the RCSJ model is the same as that the Josephson junction part.
Then, Eq.~(\ref{eq25}) actually becomes 
\begin{eqnarray}
\dot{\phi}={{4eV}\over {\hbar }}
\end{eqnarray}
due to the presence of the two gauge invariant contributions.
This indicates that if the capacitance contribution is included the observed Josephson relation is not obtained in the standard theory.

\subsection{New superconducting state case}

Now, consider the new superconducting state case. 
We only limit the consideration of the flux quantization problem, here. The ac Josephson effect is considered, later.
From Eqs.~(\ref{pr2}) and (\ref{gamma1}), and the relation $F=\hbar \gamma + \mbox{constant}$, the replacement for ${\bf p}=-i\hbar \nabla$ occurs in the following manner,
\begin{eqnarray}
-i\hbar \nabla \rightarrow -i\hbar \nabla + \hbar {\bf A}^{\rm MB}_{\Psi}
\end{eqnarray}
if the contribution of the Berry connection is included in the relevant Hamiltonian.
Further, by adding the standard replacement of the momentum operator due to the presence of the magnetic field with a vector potential ${\bf A}^{\rm em}$, the following replacement is obtained,
\begin{eqnarray}
-i\hbar \nabla + \hbar {\bf A}^{\rm MB}_{\Psi} \rightarrow -i\hbar \nabla + \hbar {\bf A}^{\rm MB}_{\Psi}
+{ e \over c}{\bf A}^{\rm em}
\end{eqnarray}

We consider the situation where the current generation is due to the velocity field arising from ${\bf A}^{\rm MB
}_{\Psi}$ and ${\bf A}^{\rm em}$.
The Meissner effect leads to the exclusion of the magnetic field from 
the interior of the superconductor, which yields
\begin{eqnarray}
0=\hbar {\bf A}^{\rm MB}_{\Psi}
+{e \over c}{\bf A}^{\rm em}=-{\hbar \over 2} \nabla \chi
+{e \over c}{\bf A}^{\rm em}
\end{eqnarray}
The flux quantization arises from the above condition.
Let us consider a ring-shaped superconductor, and take a loop $C$ that goes around the hole part of the ring through deep inside of it. Then, we have the following relation
\begin{eqnarray}
\oint_C {\bf A}^{\rm em} \cdot d{\bf r} ={{c \hbar} \over {2e }}\oint_C \nabla \chi \cdot d{\bf r}={{c h } \over {2e }}w_C[\chi]
\end{eqnarray}
where $w_C[\chi]$ is the winding number given in Eq.~(\ref{eqwindingN}).
This yields the flux quantization with the flux quantum ${{c h} \over {2e }}$. 

The appearance of $\chi$ that produces supercurrernt occurs if electrons perform spin-twisting itinerant loop motion \cite{koizumi2022}. 
 The center of the spin-twisting loop is a topological singularity of the connection where the condition in Eq.~(\ref{eqSingular}) is violated. However, the amplitude of the wave function is zero at this point, thus, the wave function does not show any anomalies. 
In the new theory,`$2e$' in ${{c h} \over {2e }}$ arises from the spinor character of the spin function for the spin-twisting itinerant loop motion.

\section{The Josephson relation and the Ambegaokar-Baratoff relation}
\label{Sec3}

In the new theory, the superconductivity is due to the presence of $\chi$.
The paired electrons are not the essential ingredient. In the standard theory, the Josephson
relation and the Ambegaokar-Baratoff relation are considered to be the manifestation
that electron-pairs are supercurrent carries \cite{Ambegaokar}; however, the new theory explains them,  differently, as will be explained below.

\subsection{The Josephson relation}
Let us now consider the current flow though the Josephson junction in the new theory.
The tunneling effect can be obtained by the first order perturbation of $H_{LR}$.
To deal with the Josephson junction problem, we use the coordinate dependent basis functions $u_{n}({\bf r}), v_{n}({\bf r})$
instead of $u_{\bf k}$ and $v_{\bf k}$ \cite{deGennes,Zhu2016}.
 Then, the electron field operators become
\begin{eqnarray}
\hat{\Psi}_{\uparrow}({\bf r},t)&=&\sum_{n} e^{-{i \over 2}\hat{\chi} ({\bf r},t)}\left( \gamma_{{n} \uparrow } u_{n}({\bf r})  -\gamma^{\dagger}_{{n} \downarrow } v^{\ast}_{n}({\bf r}) \right)
\nonumber
\\
\hat{\Psi}_{\downarrow}({\bf r},t)&=&
\sum_{n} e^{-{i \over 2}\hat{\chi} ({\bf r},t)} \left( \gamma_{{n} \downarrow } u_{n}({\bf r}) +\gamma^{\dagger}_{{n} \uparrow } v^{\ast}_{n}({\bf r}) \right)
\label{f2}
\end{eqnarray}
where the particle number conserving Bogoliubov operators satisfy
\begin{eqnarray}
\gamma_{n \sigma}|{\rm Gnd}(N) \rangle=0
\end{eqnarray}
with $N$ being the total number of particles.
The operator $e^{-{i \over 2} \hat{\chi}({\bf r},t)}$ is the coordinate and time dependent version of $e^{-{i \over 2} \hat{X}}$ appearing in Eq.~(\ref{eqNumberChanging}). The operation of $e^{-{i \over 2} \hat{\chi}({\bf r},t)}$ on the ground state yields 
\begin{eqnarray}
e^{-{i \over 2} \hat{\chi}({\bf r},t)}|{\rm Gnd}(N) \rangle= e^{-{i \over 2} {\chi}({\bf r},t)}|{\rm Gnd}(N-1) \rangle
\label{eqBC}
\end{eqnarray}
where the $U(1)$ phase factor $e^{-{i \over 2} {\chi}({\bf r},t)}$ on the right hand side. It arises from the phase change brought by the Berry connection.

We consider a lattice system where the site $i$ corresponds to the coordinate ${\bf r}_i$. Using the number-changing operators and the Bogoliubov operators defined above, annihilation and creation operators for the electrons are given by
 \begin{eqnarray}
 c_{ i \sigma} &=&\sum_{n}[ u^{n}_{i}\gamma_{n \sigma}-\sigma (v^{n}_{i})^{\ast}\gamma_{n -\sigma}^{\dagger}] e^{ -{i \over 2} \hat{\chi}_i}
 \nonumber
 \\
 c^{\dagger}_{ i \sigma} &=&\sum_{n}[ (u^{n}_{i \sigma})^{\ast}\gamma^{\dagger}_{n \sigma}-\sigma v^{n}_{i}\gamma_{n -\sigma}] e^{{i \over 2} \hat{\chi}_i}
 \label{Bog}
  \end{eqnarray}
  where $u_i^n$, $v_i^n$ , and $\hat{\chi}_i$ represent $u_n({\bf r}_i)$, $v_n({\bf r}_i)$, and $\hat{\chi}({\bf r}_i)$, respectively; values of $\sigma$ are $\sigma=1$ for up-spin state, and $\sigma=-1$ for down-spin state.
Using the particle-number conserving Bogoliubov operators, $H_{LR}$ is expressed as
  \begin{eqnarray}
H_{LR}&=&-T_{L R} e^{{ i \over 2}(\hat{\chi}_L-\hat{\chi}_R)} e^{-i {e \over {\hbar c} } \int_{{\bf r}_R}^{{\bf r}_L} d{\bf r} \cdot {\bf A}^{\rm em}}
\nonumber
\\
&& \times
\sum_{n,m} \Big[
(
(u^{n}_{L})^{\ast}\gamma^{\dagger }_{n \downarrow} + v^{n}_{L }\gamma_{n \uparrow}) ( u^{m}_{R }\gamma_{m \downarrow}+ (v^{m}_{R})^{\ast}\gamma_{m \uparrow}^{\dagger} ) 
\nonumber
\\
&&+
(
(u^{n}_{L })^{\ast}\gamma^{\dagger }_{n \uparrow} - v^{n}_{L}\gamma_{n \downarrow}) ( u^{m}_{R }\gamma_{m \uparrow}- (v^{m}_{R})^{\ast}\gamma_{m  \downarrow}^{\dagger} )  
\Big]+\mbox{h.c.}
\label{eq39}
\end{eqnarray}

Now we examine the Josephson relation. The phase corresponding to $\phi$ in Eq.~(\ref{eq:phi}) is now given by
\begin{eqnarray}
\phi_{new}=-{1 \over 2}(\chi_R-\chi_L)+{{e}\over {\hbar c}}
\int^{{\bf r}_R}_{{\bf r}_L}
 {\bf A}^{\rm em} \cdot d{\bf r}
 \label{eq:phi-new}
\end{eqnarray}
which arise from the factor $e^{{ i \over 2}(\hat{\chi}_L-\hat{\chi}_R)} e^{-i {e \over {\hbar c} } \int_{{\bf r}_R}^{{\bf r}_L}  d{\bf r} \cdot {\bf A}^{\rm em}}$ in Eq.~(\ref{eq39}).
Its time-derivative is given by
\begin{eqnarray}
\dot{\phi}_{new}&=&-{1 \over 2} \int^{{\bf r}_R}_{{\bf r}_L} \nabla \partial_t{\chi} \cdot d{\bf r}+{{e}\over {\hbar c}}
\int^{{\bf r}_R}_{{\bf r}_L} \partial_t {\bf A}^{\rm em} \cdot d{\bf r}
\nonumber
\\
&=&-{1 \over 2}\int^{{\bf r}_R}_{{\bf r}_L} \nabla \left(
\partial_t {\chi}+{{2e}\over {\hbar }}\phi^{\rm em} \right)
 \cdot d{\bf r}-{{e}\over {\hbar }}
\int^{{\bf r}_R}_{{\bf r}_L} {\bf E}^{\rm em} \cdot d{\bf r}
\nonumber
\\
&=&{{2eV} \over \hbar}
\end{eqnarray}
where relations in Eqs.~(\ref{eqV1}) and (\ref{eqV2}) are used.
By including the capacitance contribution of the junction, the correct Josephson relation is obtained in the
first order perturbation. 
In other words, the new theory obtains the correct Josephson relation with including the capacitance contribution of the Josephson junction.

\subsection{The Ambegaokar-Baratoff relation}

Let us briefly explain how the Ambegaokar-Baratoff relation is obtained in the new theory \cite{koizumi2021b}.
In order to obtain the Ambegaokar-Baratoff relation, we need to consider the case where the Bogoliubov excitations in the two superconductors are different. Then, the tunneling occurs 
as the second order perturbation. 
Let us denote the Bogoliubov operators for the left superconductor as $\gamma^{\dagger}_{L n \sigma}$ and $\gamma_{L n \sigma}$, and $\gamma^{\dagger}_{R n \sigma}$ and $\gamma_{R n \sigma}$ for the right superconductor.

The Ambegaokar-Baratoff relation is relevant near the normal-superconducting transition temperature, where the superconducting coherence length is very small.
Therefore,
it is reasonable to assume that Bogoliubov excitations in the two superconductors are different.
By using the two sets of Bogoliubov operatores, $H_{LR}$ is given by 
  \begin{eqnarray}
H_{LR}&=&-T_{L R} e^{ { i \over 2}(\hat{\chi}_L-\hat{\chi}_R)} e^{-i {e \over {\hbar c}} \int_{{\bf r}_R}^{{\bf r}_L}  d{\bf r} \cdot {\bf A}^{\rm em}}
\nonumber
\\
&& \times
\sum_{n,m} \Big[
(
(u^{n}_{L})^{\ast}\gamma^{\dagger }_{L n \downarrow} + v^{n}_{L }\gamma_{L n \uparrow}) ( u^{m}_{R }\gamma_{R m \downarrow}+ (v^{m}_{R})^{\ast}\gamma_{R m \uparrow}^{\dagger} ) 
\nonumber
\\
&&+
(
(u^{n}_{L })^{\ast}\gamma^{\dagger }_{L n \uparrow} - v^{n}_{L}\gamma_{L n \downarrow}) ( u^{m}_{R }\gamma_{R m \uparrow}- (v^{m}_{R})^{\ast}\gamma_{R m  \downarrow}^{\dagger} )  
\Big]+\mbox{h.c.}
\end{eqnarray}
The current flow across the junction is obtained by the second order perturbation of the above.
 As a consequence, the current flow is the flow of electron pairs. The standard theory considers only this case, which leads to the conclusion that the supercurrent is the flow of electron pairs.

%

The second order effective Hamiltonian with taking average over the Bogoliubov excitations is given by
\begin{eqnarray}
&&\left\langle H_{LR}{1 \over {E_0 - H_0}} H_{LR} \right\rangle
\nonumber
\\
& \approx& - \Big\langle \sum_{m, n, m', n'}T_{L R}^2 
\left[ e^{ -{i \over 2} (\hat{\chi}_L-\hat{\chi}_R)}e^{-i {e \over {\hbar c}} \int_{{\bf r}_R}^{{\bf r}_L}  d{\bf r} \cdot {\bf A}^{\rm em}}v_L^{n}u_R^{m}(\gamma_{L n \uparrow} \gamma_{R m \downarrow}-\gamma_{L n \downarrow}\gamma_{R m \uparrow})+ (L \leftrightarrow R) 
\right]
\nonumber
\\
&\times& {1 \over {\epsilon_m^{R}+\epsilon_n^{L}}}
\left[ e^{ -{i \over 2} (\hat{\chi}_L-\hat{\chi}_R)}e^{-i {e \over {\hbar c}} \int_{{\bf r}_R}^{{\bf r}_L}  d{\bf r} \cdot {\bf A}^{\rm em}}(u_{L}^{ n'}v_{R}^{ m'})^{\ast} (\gamma^{\dagger}_{L n' \downarrow} \gamma^{\dagger}_{R m' \uparrow}- \gamma^{\dagger}_{L n' \uparrow} \gamma^{\dagger}_{R m' \downarrow})+ (L \leftrightarrow R) 
\right] \Big\rangle
\nonumber
\\
&\approx& - \sum_{m, n} {{2T_{L R}^2 } \over {\epsilon_m^{R}+\epsilon_n^{L}}}
\Big[ v_L^{n}u_R^{m} (u_{L}^{ n}v_{R}^{ m})^{\ast}e^{- {i } (\hat{\chi}_L-\hat{\chi}_R)}e^{-i {{2e} \over {\hbar c}} \int_{{\bf r}_R}^{{\bf r}_L}  d{\bf r} \cdot {\bf A}^{\rm em}}
\nonumber
\\
&+&(v_L^{n}u_R^{m})^{\ast} u_{L}^{ n}v_{R}^{ m}e^{ {i} (\hat{\chi}_L-\hat{\chi}_R)}e^{i {{2e} \over {\hbar c}} \int_{{\bf r}_R}^{{\bf r}_L} d{\bf r} \cdot {\bf A}^{\rm em}}
+|u_{L}^{ n}v_{R}^{ m}|^2+ |v_{L}^{ n}u_{R}^{ m}|^2\Big]
\label{Perttransfer3}
\end{eqnarray}
From the above relation, the effective hopping Hamiltonian is obtained as
\begin{eqnarray}
H_J^{2e}=C' \cos \left( {{2e} \over {\hbar c} } \int_{{\bf r}_R}^{{\bf r}_L}  d{\bf r} \cdot \left[{\bf A}^{\rm em} +{{\hbar c} \over {2e}} \nabla \chi \right] + \alpha' \right)
\label{2e}
\end{eqnarray}
where $C'$ and $\alpha'$ are parameters given through the following relations,
\begin{eqnarray}
{ 1 \over 2} C' e^{i \alpha'}=- \sum_{m, n} {{2T_{L R}^2 } \over {\epsilon_m^{R}+\epsilon_n^{L}}}(v_L^{n}u_R^{m})^{\ast} u_{L}^{ n}v_{R}^{ m}
\end{eqnarray}
This is the well-known result in the standard theory, and gives rise to the Ambegaokar-Baratoff relation for the dc Josephson effect \cite{Ambegaokar}. The same relation is obtained in the new theory as shown above.

%

\section{A practical way to obtain ${\bf A}^{\rm MB}_{\Psi}$ in lattice systems}
\label{Sec4}

The Berry connection ${\bf A}^{\rm MB}_{\Psi}$ cannot be obtained form the Schr\"{o}dinger equation. 
We explain how to obtain ${\bf A}^{\rm MB}_{\Psi}$ in real calculations. It is obtained from  continuity equations.

\subsection{Particle-number conserving Bogoliubov-de Gennes equations}

First, we assume that the electronic Hamiltonian is given by
\begin{eqnarray}
H=\sum_{\sigma} \int d^3 r \hat{\Psi}^{\dagger}_{\sigma}({\bf r}) h({\bf r}) \hat{\Psi}_{\sigma}({\bf r}) 
-{1 \over 2} \sum_{\sigma, \sigma'}\int d^3 r d^3 r' V_{\rm eff}({\bf r}, {\bf r}') \hat{\Psi}^{\dagger}_{\sigma}({\bf r}) \hat{\Psi}^{\dagger}_{\sigma'}({\bf r}') \hat{\Psi}_{\sigma'}({\bf r}') \hat{\Psi}_{\sigma}({\bf r}) 
\nonumber
\\
\end{eqnarray}
where $h({\bf r})$ is the single-particle Hamiltonian given by
\begin{eqnarray}
h({\bf r})={ 1 \over {2m_e}} \left( { \hbar \over i} \nabla +{e \over c} {\bf A}^{\rm em} \right)^2+U({\bf r})-\mu 
\end{eqnarray}
and $-V_{\rm eff}$ is the effective interaction between electrons.
We consider the mean field solution of the above by neglecting the time-dependence of $\chi$. 
The mean-field version of the above Hamiltonian is given by
\begin{eqnarray}
H^{\rm MF}&=&\sum_{\sigma} \int d^3 r \hat{\Psi}^{\dagger}_{\sigma}({\bf r}) h({\bf r}) \hat{\Psi}_{\sigma}({\bf r}) 
+\int d^3 r d^3 r' 
\left[ \Delta({\bf r}, {\bf r}')\hat{\Psi}^{\dagger}_{\uparrow}({\bf r}) \hat{\Psi}^{\dagger}_{\downarrow}({\bf r}') e^{-{i \over 2}(\hat{\chi}({\bf r}) +\hat{\chi}({\bf r}')) }
+{\rm h. c.} \right]
\nonumber
\\
&&+\int d^3 r d^3 r' 
{ {|\Delta({\bf r}, {\bf r}')|^2} \over {V_{\rm eff}({\bf r}, {\bf r}') }}
\nonumber
\\
\label{eq43}
\end{eqnarray}
where the gap function $\Delta({\bf r}, {\bf r}')$ is given by 
\begin{eqnarray}
 \Delta({\bf r}, {\bf r}')= V_{\rm eff}({\bf r}, {\bf r}')\left\langle e^{{i \over 2}(\hat{\chi}({\bf r}) +\hat{\chi}({\bf r}')) }
\hat{\Psi}_{\uparrow}({\bf r}) \hat{\Psi}_{\downarrow} ({\bf r'}) \right\rangle
\label{eqDelta0}
\end{eqnarray}

The following commutation relations are obtained from Eqs.~(\ref{f2}), (\ref{eq43}), and the commutation relations for the Bogoliubov operators: 
\begin{eqnarray}
\left[\hat{\Psi}_{\uparrow }({\bf r}) , {H}_{\rm MF} \right]&=&
{h}({\bf r})\hat{\Psi}_{\uparrow }({\bf r})+\int d^3 r' \Delta({\bf r},{\bf r}')\hat{\Psi}^{\dagger}_{\downarrow }({\bf r}')e^{-{i \over 2}(\hat{\chi}({\bf r}) +\hat{\chi}({\bf r}')) }
\nonumber
\\
\left[\hat{\Psi}_{\downarrow }({\bf r}) , {H}_{\rm MF} \right] &=&{h}({\bf r})\hat{\Psi}_{\downarrow }({\bf r})-\int d^3 r' \Delta({\bf r},{\bf r}')\hat{\Psi}^{\dagger}_{\uparrow }({\bf r}')e^{-{i \over 2}(\hat{\chi}({\bf r}) +\hat{\chi}({\bf r}')) }
\label{deG1}
\end{eqnarray}
The particle number conserving Bogoliubov operators $\gamma_{n \sigma}$ and $\gamma^{\dagger}_{n \sigma}$ are so constructed that they satisfy
\begin{eqnarray}
\left[ {H}_{\rm MF}, \gamma_{n \sigma } \right]=-\epsilon_n \gamma_{n \sigma}, \quad \left[{H}_{\rm MF}, \gamma^{\dagger}_{n \sigma } \right] =\epsilon_n \gamma^{\dagger}_{n \sigma}
\label{deG2}
\end{eqnarray}
with $\epsilon_n \geq 0$. The condition $\epsilon_n \geq 0$ is required since the ground state is the vacuum of the Bogoliubov excitations.
By this construction, 
${H}_{\rm MF}$ is diagonalized as
\begin{eqnarray}
{H}_{\rm MF}=E_g + \sum_{n, \sigma} \epsilon_n \gamma^{\dagger}_{n \sigma}\gamma_{n \sigma}
\label{deG3}
\end{eqnarray}
where $E_g$ is the ground state energy.
From Eqs.~(\ref{deG1}), (\ref{deG2}), and (\ref{deG3}), the equations for the required Bogoliubov operators (i.e, equations to obtained parameters $u_n$'s,  $v_n$'s, and $\epsilon_n$'s for the
Bogoliubov operators) are obtained as
\begin{eqnarray}
\epsilon_n e^{-{i \over 2} \hat{\chi}({\bf r})}u_n({\bf r})&=&
h({\bf r}) e^{-{i \over 2}\hat{\chi}({\bf r})}u_n({\bf r})+\int d^3 r' \Delta ({\bf r},{\bf r}')e^{-{i \over 2}\hat{\chi}({\bf r})}v_n({\bf r}')
\nonumber
\\
\epsilon_n e^{-{i \over 2} \hat{\chi}({\bf r})}v^{\ast}_n({\bf r})&=&-
 h({\bf r}) e^{-{i \over 2} \hat{\chi}({\bf r})}v^{\ast}_n({\bf r})+\int d^3r' \Delta ({\bf r},{\bf r}')e^{-{i \over 2}\hat{\chi}({\bf r})}u^{\ast}_n({\bf r}')
\end{eqnarray}
The above are cast into the following
\begin{eqnarray}
\epsilon_n u_n({\bf r})&=&
\bar{h}({\bf r}) u_n({\bf r})+\int d^3 r'\Delta ({\bf r},{\bf r}')v_n({\bf r}')
\nonumber
\\
\epsilon_n v_n({\bf r})&=&-
 \bar{h}^{\ast}({\bf r}) v_n({\bf r})+\int d^3 r'\Delta^{\ast}({\bf r},{\bf r}')u_n({\bf r}')
 \label{e1}
\end{eqnarray}
using the relation in Eq.~(\ref{eqBC}),
where the single-particle Hamiltonian $\bar{h}$ is defined as
\begin{eqnarray}
\bar{h}({\bf r})={ 1 \over {2m_e}} \left( { \hbar \over i} \nabla +{e \over c} {\bf A}^{\rm em}-{ \hbar \over 2} \nabla \chi \right)^2+U({\bf r})-\mu 
 \label{e2}
\end{eqnarray}
and the energy gap function $\Delta({\bf r}, {\bf r}')$ is given by
\begin{eqnarray}
\Delta({\bf r}, {\bf r}')=V_{\rm eff}({\bf r}, {\bf r}')\sum_n \left[ u_n({\bf r}) v^{\ast}_n({\bf r}')\left(1- f(\epsilon_n)\right)-u_n({\bf r}') v^{\ast}_n({\bf r})f(\epsilon_n) \right]
 \label{e3}
\end{eqnarray}
Here, $f(\epsilon_n)$ is the Fermi function. 

\subsection{Calculation of $\Psi_0$}

By solving the system of equations Eqs.~(\ref{e1}), (\ref{e2}), and (\ref{e3}), self-consistently, $\Psi_0$ is obtained, first.
This is a currentless state in accordance with the so-called ``Bloch's theorem'' \cite{Bohm-Bloch}. 

\subsection{Calculation of $\chi$}

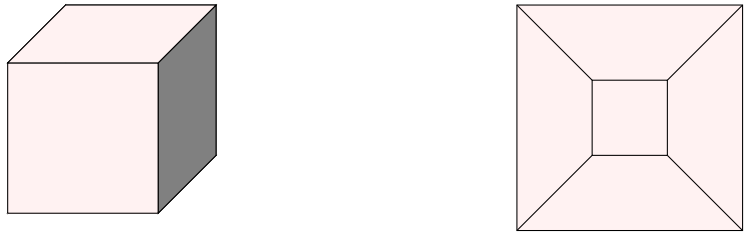
\begin{figure}[H]
  \begin{center}
  \newcommand{\Depth}{2}
\newcommand{\Height}{2}
\newcommand{\Width}{2}
\begin{tikzpicture}
\coordinate (O) at (0,0,0);
\coordinate (A) at (0,\Width,0);
\coordinate (B) at (0,\Width,\Height);
\coordinate (C) at (0,0,\Height);
\coordinate (D) at (\Depth,0,0);
\coordinate (E) at (\Depth,\Width,0);
\coordinate (F) at (\Depth,\Width,\Height);
\coordinate (G) at (\Depth,0,\Height);

\draw[fill=red!5] (O) -- (C) -- (G) -- (D) -- cycle;
\draw[fill=red!5] (O) -- (A) -- (E) -- (D) -- cycle;
\draw[fill=red!5] (O) -- (A) -- (B) -- (C) -- cycle;
\draw[fill=black!50,opacity=0.6] (D) -- (E) -- (F) -- (G) -- cycle;
\draw[fill=red!5,opacity=0.6] (C) -- (B) -- (F) -- (G) -- cycle;
\draw[fill=red!5,opacity=0.6] (A) -- (B) -- (F) -- (E) -- cycle;
\quad
\coordinate (O) at (7,0);
\coordinate (A) at (7+0.5*\Width,0);
\coordinate (B) at (7+0.5*\Width, 0.5*\Width);
\coordinate (C) at (7+0,0.5*\Width);
\coordinate (O1) at (7-0.5*\Width,-0.5*\Width);
\coordinate (A1) at (7+0.5*2*\Width,-0.5*\Width);
\coordinate (B1) at (7+0.5*2*\Width, 0.5*2*\Width);
\coordinate (C1) at (7-0.5*\Width,0.5*2*\Width);

    \draw[fill=red!5]  (O1) -- (A1) -- (B1) -- (C1) -- cycle;
  \draw (O) -- (A) -- (B) -- (C) -- cycle;
  \draw (O) -- (O1);
  \draw (A) -- (A1);
  \draw (B) -- (B1);
  \draw (C) -- (C1);
\end{tikzpicture}
  \end{center}
  \caption{Flattening a 3D lattice to 2D lattice. The shaded face in the 3D lattice is removed and converted to a 2D lattice.}
  \label{3D2D1}
\end{figure}

\begin{figure}[H]
  \begin{center}
  \newcommand{\Depth}{2}
\newcommand{\Height}{2}
\newcommand{\Width}{2}
  \begin{tikzpicture}
\coordinate (O) at (0,0,0);
\coordinate (A) at (0,\Width,0);
\coordinate (B) at (0,\Width,\Height);
\coordinate (C) at (0,0,\Height);
\coordinate (D) at (\Depth,0,0);
\coordinate (E) at (\Depth,\Width,0);
\coordinate (F) at (\Depth,\Width,\Height);
\coordinate (G) at (\Depth,0,\Height);
\coordinate (D2) at (2*\Depth,0,0);
\coordinate (E2) at (2*\Depth,\Width,0);
\coordinate (F2) at (2*\Depth,\Width,\Height);
\coordinate (G2) at (2*\Depth,0,\Height);

\draw[fill=red!5] (O) -- (C) -- (G) -- (D) -- cycle;
\draw[fill=red!5] (D) -- (G) -- (G2) -- (D2) -- cycle;
\draw[fill=red!5] (O) -- (A) -- (E) -- (D) -- cycle;
\draw[fill=red!5] (D2) -- (E2) -- (E) -- (D) -- cycle;
\draw[fill=red!5] (O) -- (A) -- (B) -- (C) -- cycle;
\draw[fill=black!50,opacity=0.6] (D) -- (E) -- (F) -- (G) -- cycle;
\draw[fill=red!5,opacity=0.6] (C) -- (B) -- (F) -- (G) -- cycle;
\draw[fill=red!5,opacity=0.6] (A) -- (B) -- (F) -- (E) -- cycle;
\draw[fill=black!50,opacity=0.6] (D2) -- (E2) -- (F2) -- (G2) -- cycle;
\draw[fill=red!5,opacity=0.6] (C) -- (B) -- (F2) -- (G2) -- cycle;
\draw[fill=red!5,opacity=0.6] (A) -- (B) -- (F2) -- (E2) -- cycle;

\coordinate (O) at (9+0,0);
\coordinate (A) at (9+0.5*\Width,0);
\coordinate (B) at (9+0.5*\Width, 0.5*\Width);
\coordinate (C) at (9+0,0.5*\Width);
\coordinate (O1) at (9-0.5*\Width,-0.5*\Width);
\coordinate (A1) at (9+0.5*2*\Width,-0.5*\Width);
\coordinate (B1) at (9+0.5*2*\Width, 0.5*2*\Width);
\coordinate (C1) at (9-0.5*\Width,0.5*2*\Width);
\coordinate (O2) at (9-0.5*2*\Width,-0.5*2*\Width);
\coordinate (A2) at (9+0.5*3*\Width,-0.5*2*\Width);
\coordinate (B2) at (9+0.5*3*\Width, 0.5*3*\Width);
\coordinate (C2) at (9-0.5*2*\Width,0.5*3*\Width);

  \draw[fill=red!5] (O2) -- (A2) -- (B2) -- (C2) -- cycle;
  \draw (O) -- (A) -- (B) -- (C) -- cycle;
  \draw(O1) -- (A1) -- (B1) -- (C1) -- cycle;
  \draw (O) -- (O1);
  \draw (A) -- (A1);
  \draw (B) -- (B1);
  \draw (C) -- (C1);
  \draw (O2) -- (O1);
  \draw (A2) -- (A1);
  \draw (B2) -- (B1);
  \draw (C2) -- (C1);
\end{tikzpicture}
  \end{center}
    \caption{Flattening a 3D lattice to 2D lattice. The two shaded faces in the 3D lattice are removed and converted to a 2D lattice.}
  \label{3D2D2}
\end{figure}
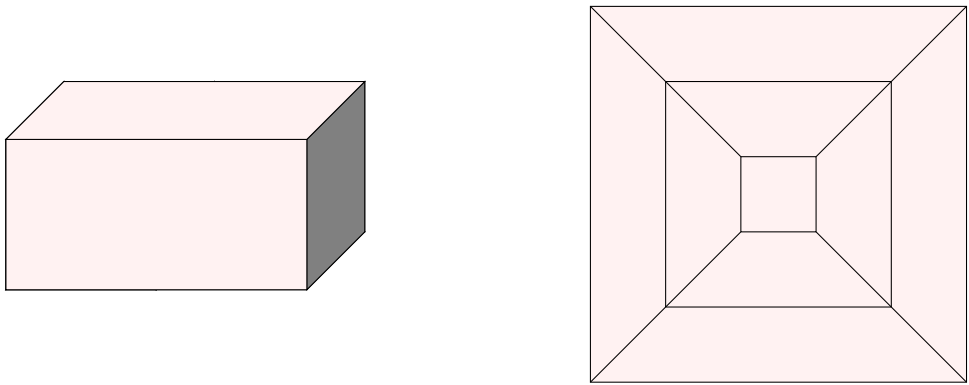

Next, $\chi$ is obtained by requiring the following two conditions, 
\begin{enumerate}

\item the single-valuedness requirement of $\Psi$ as a function of coordinates

\item the requirement of the local charge conservation

\end{enumerate}
We consider lattice systems and explain a way to obtain values of $\chi$ at lattice sites.
The value of $\chi$ at the $j$th lattice site is denoted by $\chi_j$.
For this purpose, a three dimensional lattice is flattened to a two dimensional one.
Examples are shown in Figs.~\ref{3D2D1} and \ref{3D2D2}.

We take the branch of $\chi_j$ that satisfies the difference of value between sites connected by bonds (say site $k$) is in the range,
\begin{eqnarray}
-\pi \le \chi_{j} -\chi_k < \pi
\end{eqnarray}
 
 We obtain $(\chi_k-\chi_j)$'s by minimizing the following functional 
\begin{eqnarray}
F[\nabla \chi]=E[\nabla \chi]+\sum_{\ell=1}^{N_{\rm loop}} { {\lambda_{\ell}}}\left(  \oint_{C_\ell} \nabla \chi \cdot d {\bf r}-2 \pi w_{C_{\ell}}[\chi] \right), 
\label{functional}
\end{eqnarray}
where 
\begin{eqnarray}
E[\nabla \chi]=\langle {\Psi} | H_{MF}|{\Psi} \rangle
\label{energyf}
\end{eqnarray}
Here, Lagrange multipliers $\lambda_{\ell}$'s are  introduced to impose the winding number conditions for loops; $w_{C_{\ell}}[\chi]$ is the winding number around the loop $C_{\ell}$ given by
\begin{eqnarray}
 w_{C_{\ell}}[\chi]={1 \over {2\pi}} \oint_{C_\ell} \nabla \chi \cdot d {\bf r}
 \end{eqnarray}
where
 $\{ C_1, \cdots, C_{N_{\rm loop}} \}$ are boundaries of plaques of the lattice, where $N_{\rm loop}$ is equal to the
number of plaques of the lattice. In Fig.~\ref{3D2D1}, $N_{\rm loop}=5$ case is depicted; in Fig.~\ref{3D2D2}, $N_{\rm loop}=9$ case is depicted.
The winding numbers are so chosen that the obtained wave function becomes a single-valued function of coordinates.
The values of $(\chi_k-\chi_j)$'s are obtained as solutions of the following system of equations;
\begin{eqnarray}
{{\delta E[\nabla \chi]} \over {\delta \nabla \chi}}&+&\sum_{\ell=1}^{N_{\rm loop}} { {\lambda_{\ell}}} {{\delta } \over {\delta \nabla \chi}} \oint_{C_\ell} \nabla \chi \cdot d {\bf r}=0
\label{Feq1}
\\
 \oint_{C_\ell} \nabla \chi \cdot d {\bf r}&=&2 \pi w_{C_{\ell}}[\chi]
 \label{Feq2}
\end{eqnarray}
which become 
\begin{eqnarray}
{{\partial  E( \{ \tau_{k \leftarrow j} \}) } \over {\partial \tau_{k \leftarrow j} }}&+&\sum_{\ell=1}^{N_{\rm loop}}  { {\lambda_{\ell}}} 
{{\partial } \over {\partial \tau_{k \leftarrow j}}} \sum_{k \leftarrow j} L_{k \leftarrow j}^{\ell}\tau_{k \leftarrow j} =0
\label{Feq1b}
\\
 \sum_{k \leftarrow j} L_{k \leftarrow j}^{\ell}\tau_{k \leftarrow j} &=& 2 \pi w_{C_{\ell}}[\chi]
 \label{Feq2b}
\end{eqnarray}
in the lattice system. Here, the sum is taken over the bonds ${k \leftarrow j}$, and $\tau_{ k \leftarrow j}$ is the difference of $\chi$ for the bond $\{k \leftarrow j \}$ given by
\begin{eqnarray}
\tau_{ k \leftarrow j}=\chi_k -\chi_j 
\end{eqnarray}
$L_{k \leftarrow j}^{\ell}$ denotes quantities defined below
\begin{eqnarray}
L_{k \leftarrow j}^{\ell} =
\left\{
\begin{array}{cl}
-1 & \mbox{ if  $ k \leftarrow j$ exists in $C_{\ell}$ in the clockwise direction}
\\
1 & \mbox{ if  $ k \leftarrow j$ exists in $C_{\ell}$ in the counterclockwise direction}
\\
0 & \mbox{ if  $ k \leftarrow j$ does not exist in $C_{\ell}$}
\end{array}
\right.
\nonumber
\\
\end{eqnarray}

\begin{figure}[H]
  \begin{center}
  \newcommand{\Depth}{2}
\newcommand{\Height}{2}
\newcommand{\Width}{2}
\begin{tikzpicture}
\coordinate (O) at (0,0);
\coordinate (A) at (0.5*\Width,0);
\coordinate (B) at (0.5*\Width, 0.5*\Width);
\coordinate (C) at (0,0.5*\Width);
\coordinate (O1) at (-0.5*\Width,-0.5*\Width);
\coordinate (A1) at (0.5*2*\Width,-0.5*\Width);
\coordinate (B1) at (0.5*2*\Width, 0.5*2*\Width);
\coordinate (C1) at (-0.5*\Width,0.5*2*\Width);

    \draw[fill=red!5]  (O1) -- (A1) -- (B1) -- (C1) -- cycle;
  \draw (O) -- (A) -- (B) -- (C) -- cycle;
  \draw (O) -- (O1);
  \draw (A) -- (A1);
  \draw (B) -- (B1);
  \draw (C) -- (C1);
\coordinate (O) at (7,0);
\coordinate (A) at (7+0.5*\Width,0);
\coordinate (B) at (7+0.5*\Width, 0.5*\Width);
\coordinate (C) at (7+0,0.5*\Width);
\coordinate (O1) at (7-0.5*\Width,-0.5*\Width);
\coordinate (A1) at (7+0.5*2*\Width,-0.5*\Width);
\coordinate (B1) at (7+0.5*2*\Width, 0.5*2*\Width);
\coordinate (C1) at (7-0.5*\Width,0.5*2*\Width);
   \draw[fill=red!5]  (O1) -- (A1) -- (B1) -- (C1);
  \draw (O) -- (A) -- (B) -- (C) ;
  \draw (O) -- (O1);
\end{tikzpicture}
  \end{center}
  \caption{Construction of a simply-connected lattice by removing five bonds.}
  \label{simply-conected}
\end{figure}
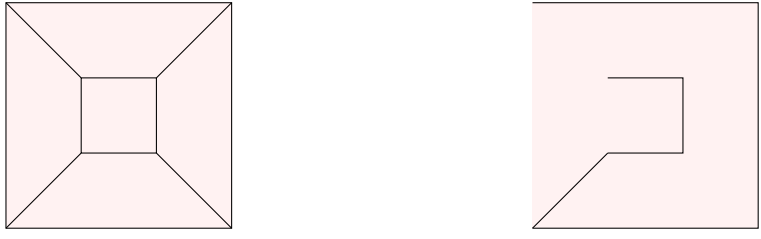

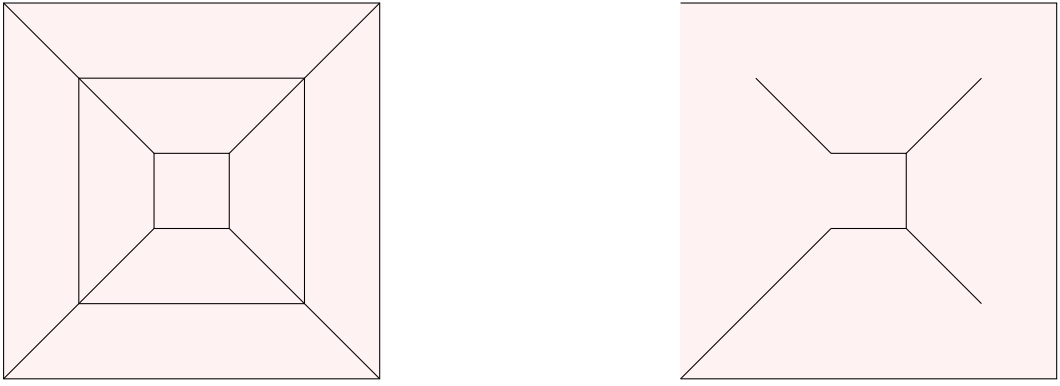
\begin{figure}[H]
  \begin{center}
  \newcommand{\Depth}{2}
\newcommand{\Height}{2}
\newcommand{\Width}{2}
  \begin{tikzpicture}
  
\coordinate (O) at (0,0);
\coordinate (A) at (0.5*\Width,0);
\coordinate (B) at (0.5*\Width, 0.5*\Width);
\coordinate (C) at (0,0.5*\Width);
\coordinate (O1) at (-0.5*\Width,-0.5*\Width);
\coordinate (A1) at (0.5*2*\Width,-0.5*\Width);
\coordinate (B1) at (0.5*2*\Width, 0.5*2*\Width);
\coordinate (C1) at (-0.5*\Width,0.5*2*\Width);
\coordinate (O2) at (-0.5*2*\Width,-0.5*2*\Width);
\coordinate (A2) at (0.5*3*\Width,-0.5*2*\Width);
\coordinate (B2) at (0.5*3*\Width, 0.5*3*\Width);
\coordinate (C2) at (-0.5*2*\Width,0.5*3*\Width);

  \draw[fill=red!5] (O2) -- (A2) -- (B2) -- (C2) -- cycle;
  \draw (O) -- (A) -- (B) -- (C) -- cycle;
  \draw(O1) -- (A1) -- (B1) -- (C1) -- cycle;
  \draw (O) -- (O1);
  \draw (A) -- (A1);
  \draw (B) -- (B1);
  \draw (C) -- (C1);
  \draw (O2) -- (O1);
  \draw (A2) -- (A1);
  \draw (B2) -- (B1);
  \draw (C2) -- (C1);

\coordinate (O) at (9+0,0);
\coordinate (A) at (9+0.5*\Width,0);
\coordinate (B) at (9+0.5*\Width, 0.5*\Width);
\coordinate (C) at (9+0,0.5*\Width);
\coordinate (O1) at (9-0.5*\Width,-0.5*\Width);
\coordinate (A1) at (9+0.5*2*\Width,-0.5*\Width);
\coordinate (B1) at (9+0.5*2*\Width, 0.5*2*\Width);
\coordinate (C1) at (9-0.5*\Width,0.5*2*\Width);
\coordinate (O2) at (9-0.5*2*\Width,-0.5*2*\Width);
\coordinate (A2) at (9+0.5*3*\Width,-0.5*2*\Width);
\coordinate (B2) at (9+0.5*3*\Width, 0.5*3*\Width);
\coordinate (C2) at (9-0.5*2*\Width,0.5*3*\Width);

  \draw[fill=red!5] (O2) -- (A2) -- (B2) -- (C2);
  \draw (O) -- (A) -- (B) -- (C);
  \draw (O) -- (O1);
  \draw (A) -- (A1);
 \draw (B) -- (B1);
 \draw (C) -- (C1);
  \draw (O2) -- (O1);
\end{tikzpicture}
  \end{center}
    \caption{Construction of a simply-connected lattice by removing nine bonds.}
  \label{simply-connected2}
\end{figure}
After $(\chi_k-\chi_j)$'s are obtained, we rebuild $\chi$ from them.
For this purpose, we remove some of the bonds from the flattened two-dimensional lattice so that
the connection of the bonds becomes a simply-connected one, i.e., there is only one path to reach a lattice point from the first lattice point denoted by `$1$'. Then, $\chi_k$ is obtained as
\begin{eqnarray}
\chi_k \approx \chi_1+ \int_{C_{1 \rightarrow k}} \nabla \chi \cdot d{\bf r}
\label{rebuilchi}
\end{eqnarray}
where $C_{1 \rightarrow k}$ is the unique path that connects from the site $1$ to the site $k$.
Note that the integration here means actually a sum.

In real calculations, we can obtain $\tau_{ j \leftarrow i}$'s without obtaining ${ {\lambda_{\ell}}}$'s. Actually, we use the conservation of the local charge.
 Let us explain this method below.
First, we note that the current through the directed bond $j \leftarrow i$ is given by
\begin{eqnarray}
J_{ j \leftarrow i}= {{2e} \over \hbar}  {{\partial E} \over {\partial \tau_{ j \leftarrow i}}}
\end{eqnarray}
Then, the conservation of charge at site $j$ is given by
\begin{eqnarray}
0=J^{\rm EX} _{ j} + \sum_i {{2e} \over \hbar}  {{\partial E} \over {\partial \tau_{ j \leftarrow i}}}
\label{Feq3}
\end{eqnarray}
where $J^{\rm EX} _{ j}$ is the external current that enters through site $j$. We use this in place of Eq.~(\ref{Feq1b}).
In order to impose conditions in Eq.~(\ref{Feq2b}), $\tau_{ j \leftarrow i}$ is split into a multi-valued part $\tau_{ j \leftarrow i}^0$ and 
single-valued part $f_{ j \leftarrow i}$ as
\begin{eqnarray}
\tau_{ j \leftarrow i}=\tau_{ j \leftarrow i}^0 + f_{ j \leftarrow i}
\end{eqnarray}
where $\tau_{ j \leftarrow i}^0$ satisfies the constraint in Eq.~(\ref{Feq2b})
\begin{eqnarray}
w_{C_{\ell}}[\chi]={ 1 \over {2\pi}} \sum_{i=1}^{N_{\ell}}  \tau^0_{{C_{\ell}(i+1)} \leftarrow {C_{\ell}(i)}}
\end{eqnarray}
and $f_{ j \leftarrow i}$ satisfies
\begin{eqnarray}
0={ 1 \over {2\pi}} \sum_{i=1}^{N_{\ell}}  f_{{C_{\ell}(i+1)} \leftarrow {C_{\ell}(i)}}
\end{eqnarray}

The number of $\tau_{ j \leftarrow i}$ to be evaluated is equal to the number of the bonds.
The number of equations in Eq.~(\ref{Feq2b}) is equal to the number of the plaques.
The number of equations from Eq.~(\ref{Feq3}) is equal to the number of sites$-1$, where $-1$ comes from the fact that the total charge is fixed in the calculation, thus, requiring local charge conservation at all sites is redundant by one.
The number of unknowns to be obtained and the number of equations have the following relation
\begin{eqnarray}
[\mbox{\# bonds}]=[\mbox{\# plaques}]+[\mbox{\# sites}-1]
\label{Euler1}
\end{eqnarray}
where `$\mbox{\# A}$' means `the number of $A$'.
It is interesting to note that this actually agrees with the Euler's theorem for the two-dimensional lattice given by
\begin{eqnarray}
[\mbox{\# edges}]=[\mbox{\# faces}]+[\mbox{\# vertices}-1]
\label{Euler2}
\end{eqnarray}
Example calculations will be found in Ref.~\cite{Koizumi2021c}.

\section{Absence of dissipative quantum phase transition in Josephson junction}
\label{Sec5}

Due to the advent of superconducting qubits, a theory for quantization of the circuit Hamiltonian for 
superconducting qubit systems has been developed \cite{Devoret1997}.
In this theory, Josephson junctions are important elements.
The new theory gives different description of them, thus, how the
modification of the theory occurs is very important. 
Actually, the theory is unaltered, practically.
However, the dissipative quantum phase transition in Josephson junction predicted by the standard theory \cite{Schmid1983,PhysRevX2021b} is absent in the new theory. 
The absence of it, however, is in agreement with the recent experimental result \cite{PhysRevX2021a,PhysRevX2021c}.
We will explain it in this section.

\subsection{Lagrangians for circuit elements}

We first, consider the quantization of circuits containing Josephson junctions \cite{Devoret1997,Nori2017}.
As usual, 
the node flux at the $n$th node is given by 
\begin{eqnarray}
\Phi_n = c \int^t_{-\infty} V_n(t') dt'
\label{eq1}
\end{eqnarray}
where $V_n$ is node voltage. It is used as the dynamical variable.
The voltage between nodes $n$ and $m$ is given by
\begin{eqnarray}
\dot{\Phi}_n -\dot{\Phi}_m
\end{eqnarray}
The Lagrangian for a capacitor with capacitance $C$ existing between node $n$ and node $m$ is given by
\begin{eqnarray}
{\cal L}_C=
{ 1 \over {2c^2}}C \left(\dot{\Phi}_n-\dot{\Phi}_m \right)^2
\end{eqnarray}

When a superconductor exists between nodes $n$ and $m$, we employ the following formula
\begin{eqnarray}
\Phi_n -\Phi_m= \int^{{\bf r}_n}_{{\bf r}_n}
{1 \over 2} \left[  {1 \over c}{\bf A}^{\rm em}({\bf r}, t) -{{\hbar  } \over {2e}} \nabla \chi({\bf r}, t) \right] \cdot d {\bf r}
\label{eq3}
\end{eqnarray}
This is different form the standard one since the factor ${1 \over 2}$ is multiplied.
The reason for the use of it is the following; according to Eq.~(\ref{eq:phi-new}),
the time-derivative of the flux difference is calculated as
\begin{eqnarray}
\dot{\Phi}_n - \dot{\Phi}_m&=& \int^{{\bf r}_n}_{{\bf r}_m}{1 \over 2}
\left[ {1 \over c} \dot{\bf A}^{\rm em}({\bf r}, t) -{{\hbar  } \over {2e}} \partial_t \nabla \chi({\bf r}, t) \right] \cdot d {\bf r}=V
\label{eq4}
\end{eqnarray}
where $V$ is the voltage across the nodes. Thus, Eq.~(\ref{eq3}) yields the correct voltage.
The current through a Josephson junction that exists between nodes $n$ and $m$ is given by
 \begin{eqnarray}
J=J_c \sin \phi
\label{eqJphi}
\end{eqnarray}
where $\phi$ is the gauge invariant phase difference between the two superconductors in the junction, and $\phi$ should satisfy the Josephson relation $\dot{\phi}={{2eV} \over {\hbar}}$.
 According to Eq.~(\ref{eq:phi-new}), such $\phi$ is given by
\begin{eqnarray}
\phi={ e \over {\hbar } }\int_{{\bf r}_n}^{{\bf r}_m} \left[{1 \over c}{\bf A}^{\rm em} -{{\hbar  } \over {2e}} \nabla \chi \right]
={ {2e} \over {\hbar c} }\left({\Phi}_m-{\Phi}_n\right)
\label{eq-phi}
\end{eqnarray}
Then, the Lagrangian for the Josephson junction ${\cal L}_J$ including the capacitance contribution 
is given by
\begin{eqnarray}
{\cal L}_J=E_J \cos { {2e} \over {\hbar c } }\left( {\Phi}_m-{\Phi}_n\right)+{C_J \over {2 c^2}}(\dot{\Phi}_n - \dot{\Phi}_m)^2
\end{eqnarray}
with $E_J={ {\hbar  J_c} \over {2e} }$.

The Lagrangian for an inductor with inductance $L$ exiting between nodes $n$ and $m$ is given by
\begin{eqnarray}
{\cal L}_L=-{ 1 \over {2L}}\left( {\Phi}_n-{\Phi}_m\right)^2
\end{eqnarray}
The minus sign here is due to the fact that it is a potential energy term with respect to the dynamical variable
${\Phi}_n$.

\subsection{Circuit quantization for a Cooper-pair box}
\label{Cooper-pairBox}

\begin{figure}[H]
  \begin{center}
    \begin{circuitikz}[american]
 \draw (0,0)   node[ground]{} node[left]{$1$} 
        to  [barrier=$E_J$,-*] (0,2) node[above]{$3$} 
     to[short] (1,2)
    to [C=$C_J$] (1,0)
      to[short,-*] (0,0);
      \draw (1,2) 
      to[short] (2,2)
        to [C=$C_g$,-*] (4,2) node[above]{$2$} 
      to[V=$V_g$] (4,0)
      to[short] (1,0);
    \end{circuitikz}
  \end{center}
  \caption{Cooper pair box. The Josephson junction is composed of components denoted as $E_J$ and $C_J$.
  $C_g$ denotes the gate capacitor, and $V_g$ supplies the gate voltage.}
  \label{CPB}
\end{figure}
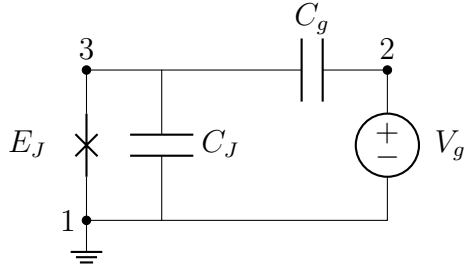

Let us consider the Cooper-pair box depicted in Fig.~\ref{CPB}.
The Lagrangian for it is given by
\begin{eqnarray}
{\cal L}_{CPB}={1 \over {2 c^2}}C_g(\dot{\Phi}_3-\dot{\Phi}_2)^2+{1 \over {2 c^2}}C_J(\dot{\Phi}_3-\dot{\Phi}_1)^2
+E_J \cos { {2e} \over {\hbar c } }\left( {\Phi}_3-{\Phi}_1\right) 
\nonumber
\\
\end{eqnarray}
The conjugate momentum for $\Phi_3$ is
\begin{eqnarray}
{ {\partial {\cal L}_{CPB} } \over {\partial \dot{\Phi}_3} }={1 \over {c^2}}C_g(\dot{\Phi}_3-\dot{\Phi}_2)+{1 \over { c^2}}C_J(\dot{\Phi}_3-\dot{\Phi}_1)
={1 \over c}Q_3
\end{eqnarray}
where $Q_3$ is charge at node $3$ given by
\begin{eqnarray}
Q_3={1 \over c}C_g(\dot{\Phi}_3-\dot{\Phi}_2)+{1 \over c}C_J(\dot{\Phi}_3-\dot{\Phi}_1)
\end{eqnarray}
The Lagrange equation for $\Phi_3$ and $\dot{\Phi}_3$ yields the equation of the conservation for the charge at node $3$,
\begin{eqnarray}
\partial_t{Q_3}+J_c \sin { {2e} \over {\hbar c} }\left( {\Phi}_3-{\Phi}_1\right)=0
\end{eqnarray}
We take $\dot{\Phi}_1=0$ since it is the ground. We also have
\begin{eqnarray}
(\dot{\Phi}_1-\dot{\Phi}_2)=c V_g
\end{eqnarray}
since we have a voltage generator between nodes $1$ and $2$.
Overall, the Lagrangian is given by
\begin{eqnarray}
{\cal L}_{CPB}={1 \over {2 c^2}}(C_g+C_J)\dot{\Phi}_3^2+{1 \over c} C_g V_g \dot{\Phi}_3+{1 \over {2 c^2}}C_g\dot{\Phi}_3^2
+E_J\cos { {2e} \over {\hbar c } } {\Phi}_3 
\end{eqnarray}
with
$
Q_3={1 \over c}(C_g+C_J)\dot{\Phi}_3+V_g C_g 
$.

Neglecting a constant term, the Hamiltonian is obtained from ${\cal L}_{CPB}$ as
\begin{eqnarray}
{\cal H}_{CPB}={1 \over {2(C_g+C_J)}} (Q_3-C_g V_g)^2
-E_J \cos { {2e} \over {\hbar c } } {\Phi_3}
\end{eqnarray}
We introduce $\hat{n}$ and $\hat{\phi}$ by
\begin{eqnarray}
\hat{n}=-{Q_3 \over {2e}}, \quad \hat{\phi}= { {2e} \over {\hbar c} }\Phi_3
\end{eqnarray}
Then, the Hamiltonian becomes
\begin{eqnarray}
{\cal H}_{CPB}=E_C (\hat{n}-n_g)^2
-E_J \cos \hat{\phi}
\label{eq58}
\end{eqnarray}
where
\begin{eqnarray}
E_C ={(2e)^2 \over {2(C_g+C_J)}} , \quad n_g={{C_g V_g} \over {2e}}
\end{eqnarray}
From the canonical quantization condition
\begin{eqnarray}
[Q_3, \Phi_3]=-i\hbar
\end{eqnarray}
we have the following relation,
\begin{eqnarray}
[\hat{n}, \hat{\phi}]=i
\label{eq32}
\end{eqnarray}
The above leads to the commutation relation,
\begin{eqnarray}
[e^{i \hat{\phi}}, \hat{n}]=e^{i \hat{\phi}}
\end{eqnarray}
Note that the above $e^{i \hat{\phi}}$ essentially corresponds to the number-changing operator in Eq.~(\ref{eqNumberChanging}) or Eq.~(\ref{eqBC}).
Strictly speaking, $\hat{\phi}$ is not a hermitian operator \cite{Phase-Angle,Fujikawa2004}
due to the fact that $\hat{\phi}$ is an operator for an angular variable.
 If we consider the eigenstates of $\hat{n}$ given by
\begin{eqnarray}
\hat{n} | n \rangle =n| n \rangle
\end{eqnarray}
where $n$ is an integer, 
the angular variable nature of $\phi$ is manifested by the relation
\begin{eqnarray}
e^{\pm i \hat{\phi}}| n \rangle =| n \pm 1 \rangle
\end{eqnarray}
An adequate quantization of ${\cal H}_{CPB}$ is obtained using the basis $\{ | n \rangle \}$ 
\begin{eqnarray}
{\cal H}_{CPB}=\sum_n 
\left[ E_C (n-n_g)^2 | n \rangle \langle n|
-{1 \over 2} E_J (| n+1 \rangle \langle n|+| n-1 \rangle \langle n|) \right]
\label{eq36}
\end{eqnarray}
Transmon qubit states are obtained as the eigenstates of the above Hamiltonian.

\subsection{Current feeding to Josephson junction and absence of dissipative quantum phase transition}

Now, we consider the absence of the dissipative quantum phase transition in Josephson junctions.
Let us consider the following Hamiltonian
\begin{eqnarray}
{\cal H}=E_C \hat{n}^2-E_J \cos \hat{\phi} -\hat{\phi} I
\label{eq66}
\end{eqnarray}
used for the prediction of the dissipative quantum phase in the standard theory. This is obtained from Eq.~(\ref{eq58}) by putting $V_g=0$ and introducing the term $-\hat{\phi} I$ to include the current bias. 
This type of current bias was introduce by Anderson \cite{Anderson64}.
It should be noted that this Hamiltonian is not self-adjoint (or hermitian) due to the fact that 
$\hat{\phi}$ is not self-adjoint \cite{Phase-Angle,Fujikawa2004}.

In the standard theory, this Hamiltonian is expressed using the representation
\begin{eqnarray}
\hat{n}=i {\partial \over {\partial \phi}}, \quad \hat{\phi}=\phi
\label{eq67}
\end{eqnarray}
Then, it becomes
a Hamiltonian for a particle in a potential energy $
-E_J \cos {\phi} -{\phi} I$ regarding ${\phi}$ as a coordinate. This potential is often called a ``washboard potential tilted by $I$''.
The above description implicitly assumes the existence of the ket vector $|\phi \rangle$ that satisfies,
\begin{eqnarray}
 \hat{\phi} |\phi \rangle=\phi|\phi \rangle
\end{eqnarray}
However, the eigenvalue $\phi$ is ill-defined due to the fact that $\hat{\phi}$ is not self-adjoint and $\phi$ is multi-valued.
Using the above representation, it has been argued that the quantum mechanical state will extend over different branches of $\phi$ (the extended-$\phi$) due to the tunneling between wells of the washboard potential. 
If the shunt resistance $R_s$ is added to the current equation, 
the time of the charge relaxation $\tau$ is given by $\tau=R_s C_J$. Then,
by denoting the quantum resistance as $R_q=h/4e^2$, it is predicted that
superconductor-insulator transition should occur around $R_q/R_s =1$ \cite{Schmid1983,PhysRevX2021b}.
Especially, $R_s \rightarrow \infty$ leads to $I \rightarrow 0$, thus, 
the $I=0$ insulator state should be observed.
However, ${\cal H}$ with $I=0$ is essentially equivalent to ${\cal H}_{CPB}$; then, 
the system should be superconducting that gives rise to qubit states. 
Actually, the recent state-of-the-art experiment indicates that the description by  $\phi$ in a single branch (the compact-$\phi$) is appropriate, and the dissipative quantum phase transition does not happen
\cite{PhysRevX2021a,PhysRevX2021c}, contradicting the standard theory predictions.

The above discrepancy is resolved if we note that Eq.~(\ref{eq67}) is not the only 
representation that satisfies Eq.~(\ref{eq32}), and the current bias is taken into account 
as a supplementary condition.
The condition is given by 
\begin{eqnarray}
[\hat{n}, {\tilde{\phi}}]=i,  \quad \hat{\phi}={\tilde{\phi}}+\phi_0
\label{eqnphi}
\end{eqnarray}
where $\phi_0$ is the expectation value $\langle \hat{\phi} \rangle$, and ${\tilde{\phi}}$
is the fluctuation from the average that satisfies
\begin{eqnarray}
\langle {\tilde{\phi}} \rangle=0
\label{eq70}
\end{eqnarray}
The value of 
$\phi_0$ is chosen to satisfy the local charge conservation, and the current bias is given by
\begin{eqnarray}
I=J_c \sin {\phi}_0 
\end{eqnarray}
Thus, the same problem is solved using the modified version of ${\cal H}_{CPB}$ given by
\begin{eqnarray}
{\cal H}^m_{CPB}=E_C \hat{n}^2-E_J \cos (\phi_0+\tilde{\phi})
\end{eqnarray}
The term $-I\hat{\phi}$ is absent, making this Hamiltonian self-adjoint. When the system is approached to zero current bias condition $I \rightarrow 0$, the phase $\phi_0$ shifts to ${\phi}_0 \rightarrow 0$. In this way, 
the system is superconducting all the way without quantum phase transition.
Using the condition in Eq.~(\ref{eq70}), ${\cal H}^m_{CPB}$ becomes
\begin{eqnarray}
{\cal H}^m_{CPB} &=& E_C \hat{n}^2-E_J [ \cos \phi_0 \cos\tilde{\phi}-\sin \phi_0 \sin\tilde{\phi}]
\nonumber
\\
&\approx& E_C \hat{n}^2-E_J \cos \phi_0 \cos\tilde{\phi}
\end{eqnarray}
where the following approximation is used
\begin{eqnarray}
\sin\tilde{\phi} \approx \sin \langle \tilde{\phi} \rangle =0
\end{eqnarray}
The current bias reduces the barrier height from $E_J$ to $E_J \cos \phi_0$ in a similar manner as the washboard potential does. This will explain the
experimental results that are explained using the non-hermitian Hamiltonian in Eq.~(\ref{eq66}) \cite{Clarke1988,Martinis2002}.

\section{Concluding remarks}
\label{Sec6}

%

In this work, theory of superconductivity is reformulated by
 including the $U(1)$ phase neglected by Dirac. 
 The present work indicates that the wave function in the Schr\"{o}dinger representation of quantum mechanics can work as
an operator that generates topological defects and gives rise to a gauge field. Such
defects can be detected using the Berry connection.
The gauge field arising from them gives rise to a collective mode (given by $\chi$), and the supercurrent is generated by $\chi$ when the
condition in Eq.~(\ref{eq15}) is established.

The summary of the comparison between the standard and new theories is tabulated in Table~\ref{table1}.
The current by $\chi$ mode and associated number fluctuation are the key ingredients for realizing 
superconducting states. The particle number is conserved in the new theory.
The mass in the London moment is the free electron mass, and the dissipative quantum phase transition does not occur in the new theory in agreement with the experiments.
The capacitance contribution is missing in the original derivation for the Josephson relation
$\dot{\phi}={{2eV} \over \hbar} $, and when it is included, the standard theory actually disagrees with the experiment.

\begin{landscape}
 \begin{table}
  \begin{tabular}{|c||c|c|}
   \hline
 Item  & Standard Theory & New Theory \\
   \hline \hline
 Collective mode for supercurrent & Nambu-Goldstone mode &  Neglected $U(1)$ phase\\
 \hline
 Particle number conservation &  No &  Yes \\
 \hline
 $U(1)$ gauge symmetry breaking & Yes &  No \\
 \hline
Bogoliubov operators &  Particle number non-conserving  &  Particle number conserving   \\
 \hline
 Electron mass in London moment &  effective mass & free electron mass  \\
 \hline
  $\phi$ of $J=J_c \sin \phi$ in Eq.~(\ref{eqJphi}) & {\scriptsize $\theta_R-\theta_L+{{2e}\over {\hbar c}}
\int^{{\bf r}_R}_{{\bf r}_L}
 {\bf A}^{\rm em} \cdot d{\bf r}$ } in Eq.~(\ref{eq:phi}) & 
 {\scriptsize ${{\chi_L-\chi_R} \over 2}+{{e}\over {\hbar c}}
\int^{{\bf r}_R}_{{\bf r}_L}
 {\bf A}^{\rm em} \cdot d{\bf r}$ } 
 in Eq.~(\ref{eq:phi-new}) \\
  \hline
 $\dot{\phi}$ without capacitance contribution &  ${{2eV} \over \hbar} $ & ${{eV} \over \hbar} $  \\
 \hline
  $\dot{\phi}$ with capacitance contribution &  ${{4eV} \over \hbar} $ & ${{2eV} \over \hbar} $  \\
  \hline
 $[\hat{n}, \hat{\phi}]=i$ in Eq.~(\ref{eq32})
&
  $\hat{n}=i {\partial \over {\partial \phi}}, \quad \hat{\phi}=\phi$ in Eq.~
(\ref{eq67}) & 
$[\hat{n}, {\tilde{\phi}}]=i,  \quad \hat{\phi}={\tilde{\phi}}+\phi_0$ in Eq.~(\ref{eqnphi})
\\
   \hline
   Dissipative quantum phase transition & occurs & does not occur
   \\
   \hline
  \end{tabular}
  \caption{The summary of the comparison between the standard theory and new one.}
 \label{table1}
\end{table}
\end{landscape}

From the view point of the new theory, the BCS theory is the one that effectively takes into account the effect of $\chi$ by employing the particle number non-conserving ket vector. The existence of $\chi$ generates the pair potential in Eqs.~(\ref{eqDelta0}) and (\ref{e3}) with keeping the particle number fixed in the new theory; however, it is achieved by the fluctuation of the number of particles in the BCS theory.
The superconducting state is realized when the energy gap (or the pair potential) by the
electron-pairing appears near the Fermi level of the metallic normal state in the BCS theory.
 The energy gap formation and the stabilization of $\chi$ occur, simultaneously, in many superconductors; thus, $T_c$ is given by the pairing energy gap formation temperature. In this case, the pairing energy gap can be used as the superconducting order parameter.

According to the new theory, however, the pairing energy gap formation requires the presence of $\chi$. This $\chi$ is most likely generated by the spin-twisting itinerant motion of electrons, and such a motion may be realized as a circular motion of band electrons around a section of the Fermi surface. Then, its stabilization is achieved by the pairing energy gap formation explained in the BCS theory.
However, the BCS theory does not work for the cuprate superconductors; this is probably due to the fact that the stabilization of $\chi$ occur under the influence of several factors other than the electron-pair formation.
First of all, we expect the spin-twisting itinerant motion of electrons is not a circular motion of band electrons, but a circular motion of electrons around doped holes.
Since the doped holes form small polarons, this small polaron formation is an important factor.
The spin-vortex is expected to be created around each of them, thus, the spin-vortex formation is another important factor.
The pairing energy gap is most likely created by the Coulomb repulsion, which also plays an important role in the appearance of spin moments to generate spin-vortices.  Thus, the correlation effect by the Coulomb repulsion is also an important factor.
Finally, the stabilization of loop currents generated by $\chi$ is also  an important factor. 
Over all, the quantitative understanding of the cuprate superconductivity will requires the inclusion of all those factors.

%

We may apply the present new theory to other superfluid phenomena.
For superfluid phase of boson systems,
the order parameter is often defined using the Bose-Einstein condensation criterion by Penrose and Onsager \cite{Penrose56}. 
This criterion  uses the single-particle density matrix given by
\begin{eqnarray}
\rho_1({\bf r},{\bf r}',t)=\langle \hat{\Psi}_B^\dagger({\bf r},t)\hat{\Psi}_B({\bf r}',t)\rangle
\end{eqnarray}
where $\hat{\Psi}_B$ is the boson field operator.
It is reexpressed using the natural orbital basis $\{ f_i \}$ and the occupation number $n_i$ for the state $f_i$
\begin{eqnarray}
\rho_1({\bf r},{\bf r}',t)&=&\sum_i n_i(t) f_i^\ast({\bf r},t)f_i({\bf r}',t)
\nonumber
\\
&=&\Psi^\ast({\bf r},t)\Psi({\bf r}',t)+\sum_{i\neq 0} n_i(t) f_i^\ast({\bf r},t) f_i({\bf r}',t)
\end{eqnarray}
where $f_0$ is the condensate wave function with $n_0$ in the order of the total number of particles. The order parameter $\Psi({\bf r},t)$ is identified as $f_0({\bf r},t)$ \cite{LeggettBook}.

There is another definition of the order parameter given by
\begin{eqnarray}
\Psi({\bf r},t)= \langle \hat{\Psi}_B({\bf r}',t)\rangle
\label{eq105}
\end{eqnarray}
that requires the particle-number non-conserving formalism \cite{Anderson66}.
We may regard this formalism as an approximation to the one that includes the Dirac's neglected phase.
When the Dirac's neglected phase is included, a collective mode $\chi_B$ (just like $\chi$) appears, and we may take it as the superfluid current mode.
Using this mode, we define a new 
 single-particle density matrix given by
 \begin{eqnarray}
\tilde{\rho}_1({\bf r},{\bf r}',t)=\langle \hat{\rho}_B^{1/2}({\bf r},t)e^{i \hat{\chi}_B({\bf r},t)} e^{-i \hat{\chi}_B({\bf r}',t)}\hat{\rho}_B^{1/2}({\bf r}',t)\rangle
\end{eqnarray}
where $\hat{\rho}_B$ is the number-density of the particle participating in the collective mode,
and $e^{\pm i \hat{\chi}_B}$ are the number-changing operators for particles participating in the collective mode.
If we approximate it as
\begin{eqnarray}
\tilde{\rho}_1({\bf r},{\bf r}',t)
&\approx&
\left\langle \hat{\rho}_B^{1/2}({\bf r},t) e^{i \hat{\chi}_B({\bf r},t)} e^{-i \hat{\chi}_B({\bf r}',t)} \right\rangle \left\langle \hat{\rho}_B^{1/2}({\bf r}',t)\right\rangle
\nonumber
\\
&\approx&
\left\langle \hat{\rho}_B^{1/2}({\bf r},t) \right\rangle e^{i {\chi}_B({\bf r},t)} e^{-i {\chi}_B({\bf r}',t)} \left\langle \hat{\rho}_B^{1/2}({\bf r}',t)\right\rangle
\nonumber
\\
&=&
\rho_B^{1/2}({\bf r},t) e^{i {\chi}_B({\bf r},t)} e^{-i {\chi}_B({\bf r}',t)} {\rho}_B^{1/2}({\bf r}',t)
\end{eqnarray}
we will have the following order parameter
\begin{eqnarray}
\Psi({\bf r},t)=e^{-i {\chi}_B({\bf r},t)} {\rho}_B^{1/2}({\bf r},t)
\label{eq111}
\end{eqnarray}
In this way, the order parameter looks similar to Eq.~(\ref{eq105}) is obtained.
At present, we do not know the exact relation between the appearance of the order parameter in Eq.~(\ref{eq111}) and the Bose-Einstein condensation. However, a close relation is plausible,
in which the particles
in the condensate state will play a similar role as the electrons near the Fermi level play.
The reduction of the total energy is expected to occur by exploiting the interaction energy 
and the number fluctuation of the particles participating in the collective mode $\chi_B$. 

In conclusion, we present a new theory of superconductivity that includes
the neglected $U(1)$ phase by Dirac.
This theory captures major experimental results explained by the BCS theory,
and lifts shortcomings of it.

\section*{References}

\providecommand{\newblock}{}


\end{document}